\documentclass[twocolumn]{aastex701}

\DeclareRobustCommand{\okina}{%
  \raisebox{\dimexpr\fontcharht\font`A-\height}{%
    \scalebox{0.8}{`}%
  }%
}

\begin{document}

\title{Extreme Stellar Death and Galaxy Feedback at $z=2$: A Rest-frame Ultraviolet Characterization of the Strongly Lensed Superluminous Supernova 2025wny}

\author[0000-0003-3953-9532]{Willem~B.~Hoogendam}
\altaffiliation{NSF Graduate Research Fellow}
\affiliation{Institute for Astronomy, University of Hawai\okina i, 
2680 Woodlawn Drive, Honolulu, HI 96822, USA}
\email[show]{willemh@hawaii.edu}  

\author[0000-0002-6230-0151]{David~O.~Jones}
\affiliation{Institute for Astronomy, University of Hawai\okina i, 640 N. A\okina ohoku Pl., Hilo, HI 96720, USA}
\email{dojones@hawaii.edu}

\author[0000-0002-0385-0014]{Christopher~J.~Storfer}
\affiliation{Institute for Astronomy, University of Hawai\okina i, 
2680 Woodlawn Drive, Honolulu, HI 96822, USA}
\email{cstorfer@hawaii.edu}

\author[0000-0001-9695-8472]{Luca~Izzo}
\affiliation{DARK, Niels Bohr Institute, University of Copenhagen, Jagtvej 155A, 2200 Copenhagen, Denmark}
\affiliation{INAF, Osservatorio Astronomico di Capodimonte, Salita Moiariello 16, I-80131 Naples, Italy}
\email{luca.izzo@nbi.ku.dk,luca.izzo@inaf.it}

\author[0000-0002-5680-4660]{Kyle~W.~Davis}
\affiliation{Department of Astronomy and Astrophysics, University of California, Santa Cruz, CA 95064, USA}
\email{kywdavis@ucsc.edu}

\author[0009-0005-5121-2884]{Miranda~Y.~Kong}
\affiliation{Institute for Astronomy, University of Hawai\okina i, 
2680 Woodlawn Drive, Honolulu, HI 96822, USA}
\email{ykong2@hawaii.edu}

\author[0000-0002-4449-9152]{Katie~Auchettl}
\affiliation{School of Physics, The University of Melbourne, VIC 3010, Australia}
\email{katie.auchettl@unimelb.edu.au}

\author[0009-0008-1965-9012]{Aadya~Agrawal}
\affiliation{Department of Astronomy, University of Illinois at Urbana-Champaign, 1002 W. Green St., IL 61801, USA}
\email{aadyaa2@illinois.edu}

\author[0000-0002-3874-2769]{David~R.~Aguilera-Dena}
\affiliation{Department of Astronomy and Astrophysics, University of California, Santa Cruz, CA 95064, USA}
\affiliation{Instituto de Astronom\'ia, Universidad Nacional Aut\'onoma de M\'exico, Ap. 70-264, 04510 CDMX, M\'exico}
\email{daguilera@astro.unam.mx}

\author[0000-0001-6965-7789]{Kenneth~C.~Chambers}
\affiliation{Institute for Astronomy, University of Hawai\okina i, 
2680 Woodlawn Drive, Honolulu, HI 96822, USA}
\email{kchamber@hawaii.edu}

\author[0009-0007-7564-4408]{Siyuan~Chen}
\affiliation{Center for Astrophysics \textbar \, Harvard \& Smithsonian, 60 Garden Street, Cambridge, MA 02138, USA}
\email{siyuan.chen@cfa.harvard.edu}

\author[0000-0001-5486-2747]{Thomas~de~Boer}
\affiliation{Institute for Astronomy, University of Hawai\okina i, 
2680 Woodlawn Drive, Honolulu, HI 96822, USA}
\email{tdeboer@hawaii.edu}

\author[orcid=0000-0002-2445-5275,sname='Foley']{Ryan~J.~Foley}
\affiliation{Department of Astronomy and Astrophysics, University of California, Santa Cruz, CA 95064, USA}
\email{foley@ucsc.edu}

\author[0000-0002-6741-983X]{Matthew~Grayling}
\affiliation{Institute of Astronomy and Kavli Institute for Cosmology, University of Cambridge, Cambridge CB3 0HA, UK}
\email{mg2102@cam.ac.uk}

\author[orcid=0000-0001-9668-2920,sname='Hinkle']{Jason~T.~Hinkle}
\altaffiliation{NHFP Einstein Fellow}
\affiliation{Department of Astronomy, University of Illinois at Urbana-Champaign, 1002 W. Green St., IL 61801, USA}
\affiliation{NSF-Simons SkAI Institute, 172 E. Chestnut St., Suite 3500, Chicago, IL 60611, USA}
\email{jhinkle6@illinois.edu}

\author[0000-0002-4571-2306]{Jens~Hjorth}
\affiliation{DARK, Niels Bohr Institute, University of Copenhagen, Jagtvej 155A, 2200 Copenhagen, Denmark}
\email{jens@nbi.ku.dk}

\author[0000-0003-1059-9603]{Mark~E.~Huber}
\affiliation{Institute for Astronomy, University of Hawai\okina i, 
2680 Woodlawn Drive, Honolulu, HI 96822, USA}
\email{mehuber7@hawaii.edu}

\author[0000-0002-7272-5129]{Chien-Cheng~Lin}
\affiliation{Institute for Astronomy, University of Hawai\okina i, 
2680 Woodlawn Drive, Honolulu, HI 96822, USA}
\email{cclin33@hawaii.edu}

\author[0000-0002-9438-3617]{Thomas~B.~Lowe}
\affiliation{Institute for Astronomy, University of Hawai\okina i, 
2680 Woodlawn Drive, Honolulu, HI 96822, USA}
\email{tlowe@hawaii.edu}

\author[0000-0002-7965-2815]{Eugene~A.~Magnier}
\affiliation{Institute for Astronomy, University of Hawai\okina i, 
2680 Woodlawn Drive, Honolulu, HI 96822, USA}
\email{magnier@hawaii.edu}

\author[0000-0001-9846-4417]{Kaisey~S.~Mandel}
\affiliation{Institute of Astronomy and Kavli Institute for Cosmology, University of Cambridge, Cambridge CB3 0HA, UK}
\email{kmandel@ast.cam.ac.uk}

\author[0009-0006-5214-0736]{C.~Tanner~Murphey}
\affiliation{Department of Astronomy, University of Illinois at Urbana-Champaign, 1002 W. Green St., IL 61801, USA}
\affiliation{Center for Astrophysical Surveys, National Center for Supercomputing Applications, Urbana, IL 61801, USA}
\email{murphey2@illinois.edu}

\author[0000-0002-2260-4628]{Gautham~Narayan}
\affiliation{Department of Astronomy, University of Illinois at Urbana-Champaign, 1002 W. Green St., IL 61801, USA}
\affiliation{Center for Astrophysical Surveys, National Center for Supercomputing Applications, Urbana, IL 61801, USA}
\affiliation{NSF-Simons SkAI Institute, 172 E. Chestnut St., Suite 3500, Chicago, IL 60611, USA}
\email{gsn@illinois.edu}

\author[0000-0002-6639-6533]{Gregory~S.~H.~Paek}
\affiliation{Institute for Astronomy, University of Hawai\okina i, 
2680 Woodlawn Drive, Honolulu, HI 96822, USA}
\email{gregorypaek94@gmail.com}

\author[0000-0002-1092-6806]{Kishore~C.~Patra}
\affiliation{Department of Astronomy and Astrophysics, University of California, Santa Cruz, CA 95064, USA}
\email{kcpatra@ucsc.edu}

\author[0009-0000-5561-9116]{Haille~M.~L.~Perkins}
\affiliation{Department of Astronomy, University of Illinois at Urbana-Champaign, 1002 W. Green St., IL 61801, USA}
\affiliation{Center for Astrophysical Surveys, National Center for Supercomputing Applications, Urbana, IL 61801, USA}
\email{haillep2@illinois.edu}

\author[0000-0002-4410-5387]{Armin~Rest}
\affiliation{Department of Physics and Astronomy, The Johns Hopkins University, Baltimore, MD 21218, USA}
\affiliation{Space Telescope Science Institute, Baltimore, MD 21218, USA}
\email{arest@stsci.edu}

\author[0000-0003-4631-1149]{Benjamin~J.~Shappee}
\affiliation{Institute for Astronomy, University of Hawai\okina i, 
2680 Woodlawn Drive, Honolulu, HI 96822, USA}
\email{shappee@hawaii.edu}

\author[0000-0002-1341-0952]{Richard~J.~Wainscoat}
\affiliation{Institute for Astronomy, University of Hawai\okina i, 
2680 Woodlawn Drive, Honolulu, HI 96822, USA}
\email{rjw@hawaii.edu}

\begin{abstract}
We present photometric and spectroscopic follow-up observations of the strongly-lensed H-poor superluminous supernova (SLSN) 2025wny at $z = 2.0155$. We use integral-field Keck-II/KCWI spectroscopy to obtain rest-frame ultraviolet spectra spanning $11$ to $61$ days after maximum light. We use Pan-STARRS photometry to constrain the brightest image's magnification, finding $\mu_A=19.4\pm1.6$, a factor of $\sim$2--4 higher than mass-modeling estimates. We fit the Pan-STARRS photometry with a magnetar energy-injection model including $^{56}$Ni decay using the Modular Open Source Fitter for Transients (\texttt{MOSFiT}); the model can reproduce the near-maximum light curve with parameters typical of local universe SLSNe-I, including a magnetar spin period $P_{\rm spin}=4.1^{+0.7}_{-1.0}~\mathrm{ms}$ and magnetic field $B_{\perp}=2.4^{+1.1}_{-0.8}\times10^{14}~\mathrm{G}$; however, the late-time plateau suggests an additional power source.  Spectroscopically, iron-group element line blanketing is virtually absent in the rest-frame UV spectra, and the absorption features match previously identified SLSN-I features from abundance tomography.  Based on narrow ISM line metallicity estimates, SN~2025wny exploded in a sub-solar-metallicity galaxy. From host-galaxy lines in our rest-frame UV-to-optical spectrum we estimate high-velocity outflows of $\sim$575~${\rm km~s^{-1}}$ and an exceptionally high Ly$\alpha$ escape fraction of $f_{esc}^{Ly\alpha}=0.23\pm0.03$. SN~2025wny matches lower-redshift SLSNe reasonably well. Upcoming surveys like LSST and \emph{Roman} will discover additional high-redshift SLSNe, further testing whether SLSNe-I explosions remain consistent at the population level in the early universe.
\end{abstract}

\keywords{\uat{Supernovae}{1668} --- \uat{Gravitational Lensing}{670}} 

\section{Introduction} 
 Superluminous supernovae (SLSNe) are a rare class of extreme supernova explosions \citep{Quimby07_SLSN, quimby2011, Gal-Yam12, Howell17, Gal-Yam19_ARAA}. Their peak luminosities exceed those of thermonuclear and core-collapse supernovae by one to two orders of magnitude \citep[e.g.,][]{BranchWheeler_SN_Explosions_Book, Gal-Yam19_ARAA}. SLSNe evolve more slowly than typical supernovae, sometimes undulating, as they decline over months rather than weeks \citep{Nicholl17, Gal-Yam19_ARAA, Chen23_I, Chen23_II, Farah26}. Spectroscopically, SLSNe follow the typical ``type I'' and ``type II'' convention \citep{Gal-Yam19_ARAA}, which separates objects by the absence or presence of H features in their spectra \citep{Minkowski42}. 

For SLSNe-II, their similarity to interacting SNe~IIn \citep{Smartt09_ARAA, Gal-Yam17} suggests that interaction with large amounts of circumstellar material (CSM) may explain their overluminous nature \citep[e.g.,][]{Inserra18_SLSN-II}. On the other hand, there is less consensus about the nature of the additional energy source(s) for SLSNe-I, although there is growing convergence \citep[e.g.,][]{Nicholl17, gomez2024, Hinkle25_SwiftFixII} toward central engine models where the core collapse of a massive star results in energy injection from the spin-down of a newly formed magnetar \citep[e.g.,][]{Kasen10b, Woosley10}. 

Many scenarios have been proposed to explain the origin and properties of SLSNe-I, but the specific stellar evolutionary channels that lead to SLSNe and the mechanisms that power their light curves remain debated. The radioactive decay of $^{56}\mathrm{Ni}$ alone cannot fully account for the extreme luminosities of SLSNe \citep[e.g.,][]{Quimby07_SLSN, quimby2011, Chomiuk11_SLSN}, so they require additional energy from at least one other source. Beyond the magnetar model, possible additional flux contributions may come from CSM interaction with an H-poor CSM \citep[e.g.,][]{Chevalier11, Moriya13}, which \citet{Chen23_II} find a sizable fraction ($62\%$) of SLSNe-I require. The mechanism that produces this CSM may also differ between homogeneous evolution \citep[e.g.,][]{Aguilera-Dena18, Aguilera-Dena20} and pulsational pair instability \citep[e.g.,][]{Vigna-Gomez19} models. Beyond CSM, a central engine powered instead by fallback accretion \citep[e.g.,][]{Dexter13, Moriya18_fallback}, or a pair-instability explosion which produces extreme amounts of $^{56}\mathrm{Ni}$ exceeding that produced in typical SN \citep[e.g.,][]{Barkat67, Kasen11, Woosley17, Schulze24}.

 Further complicating comparisons between models and data, many models predict similar observables \citep[e.g.,][]{Sukhbold16}, and most SLSNe-I fit well with either a magnetar spin-down model or a CSM+$^{56}\mathrm{Ni}$ model. Light-curve undulations pose a serious challenge to magnetar models \citep[e.g.,][]{Nicholl17} and pure $^{56}\mathrm{Ni}$ decay, but recent work suggests that CSM interaction \citep{Chen23_II} and/or Lense-Thirring precession of a misaligned infalling accretion disk around the magnetar \citep{Farah26} may account for them. Additionally, while SLSNe-I show blue continua and strong \ion{O}{2} absorption features \citep[e.g.,][]{Quimby2018, Gal-Yam19, aamer2025}, the general lack of other strong features in their optical spectra limits the inferences possible from photospheric modeling \citep{Dessart16, Woosley21}. Finally, a possible intermediate class of ``luminous'' SNe, perhaps arising from variations in physical properties such as $^{56}\mathrm{Ni}$ or magnetar engine strength \citep{Gomez22}, further complicates efforts to provide an all-encompassing physical model. 

SLSNe-I tend to occur in low-mass, low-metallicity host galaxies with higher specific star formation rates \citep[e.g.,][]{Chen2013, Lunnan13, Lunnan14, Lunnan15, Angus2016, Perley2016, Schulze2018, Hsu2024}, possibly resulting from galaxy merger activity \citep{Orum20}, but there are some notable exceptions \citep[e.g.,][]{Bose18}. This aligns with the magnetar model: lower progenitor metallicity leads to weaker pre-supernova mass loss, preserving the angular momentum needed to form a magnetar \citep[e.g.,][]{Yoon2005_MassiveStars, Woosley2006, Aguilera-Dena18}. Other metallicity-dependent effects, like rotation, mass loss, and binarity, may lead to asymmetric CSM density distributions. Recent work has confirmed that metallicity is the most constraining factor on SLSN production \citep{Cleland23}. Taken together, SLSNe-I appear to be linked to low host-galaxy metallicities, implying that SLSNe-I arise from massive stars forming in low-metallicity environments, and their rate may increase substantially at high redshift.

Rest-frame ultraviolet (UV) data are particularly valuable for these objects, as the UV is sensitive to the iron-group element abundances set by pre-explosion stellar metallicity, nuclear processing, mixing, and/or explosion energy input. \citet{mazzali2016} find that SLSN-I are generally line-rich in the UV, showing many high-ionization lines from C, O, Si, Mg, Fe, and Ti, and a significant influence of non-thermal processes on the \ion{O}{2} and \ion{He}{1} features. Alternatively, the \emph{HST} rest-frame UV spectra of Gaia16apd have fewer absorption features and may imply a low-metallicity progenitor \citep{yan2017} and/or a central engine with a shorter spin period, lower mass, and weaker magnetic field \citep{Nicholl17_Gaia16apd}. However, UV data are sparser than optical data, and only a few SLSNe have high-quality rest-frame UV spectra \citep[see, e.g.,][]{Yan18}. 

Strongly lensed supernovae offer a unique vantage point for studying rest-frame UV emission processes that are otherwise too faint or too observationally challenging to study\footnote{For the challenges of studying high-redshift core-collapse~SNe without assistance from strong lensing, see, e.g., \citet{Siebert24}, \citet{Aparicio_2026}, \citet{Coulter26a}, and \citet{Coulter26b}.}. A second advantage conferred by strong lensing is the possibility of multiple images of the same source, potentially accompanied by time delays \citep[e.g.,][]{refsdal1964, treu2010}. These supernovae are rare, with only six such supernovae from galaxy-scale lensing systems: PS1-10afx \citep{Quimby13, quimby2014}, iPTF16geu \citep{goobar20171, more2017, dhawan2020, Gall24, Baltasar2026}, SN~2022qmx \citep{goobar2023, pierel2023, Larison2025}, and SN~2026ngr \citep{Liu26_AstroNote}, all of which were type Ia supernovae, the recent type II SN~2025mkn \citep{Lemon26}, and now, SN~2025wny \citep{Taubenberger25, Johansson25, Storfer26, Ecker26}, the first galaxy-scale strongly lensed SLSN and the subject of this manuscript. During the final preparation of this manuscript, several complementary studies of SN~2025wny were also independently submitted  \citep{Goobar26_2025wny, Li26_2025wny, Johansson26_2025wny_b, Townsend26_2025wny, Mortsell26_2025wny, Hjortlund26_25wny}. 

In this manuscript, we present new photometric and spectroscopic observations of SN~2025wny, including a comprehensive FUV spectral time series of an SLSN-I, with the latest epoch more than two months after maximum light in the rest frame. This manuscript is organized as follows: \S\ref{sec:data} describes the new observations presented in this work, \S\ref{sec:phot_analysis} presents photometric analysis of the SN properties, including physical modeling of the light curve, and \S\ref{sec:spec_analysis} details complementary spectroscopic analysis of SN~2025wny. The host-galaxy analysis is in \S\ref{sec:spec_analysis_host}. We summarize our findings in \S\ref{sec:conclusions}. Throughout this work, we adopt a flat $\Lambda$CDM cosmology with $H_0=70$~km~s$^{-1}$~Mpc$^{-1}$ and $\Omega_m=0.30$. The Milky-Way extinction is $E(B-V)=0.053$ \citep{schlegel19981, Schlafly11}.

\section{Data}\label{sec:data}

The Gravitational wave Optical Transient Observer (GOTO; \citealp{goto1, goto2}) discovered SN~2025wny on 2025-09-01 (JD 2460919.7) with the internal designation GOTO25gqt \citep{2025wny_discovery}. Subsequent follow-up observations revealed multiple images \citep{2025wny_classifcation_wise} and an SLSN spectroscopic classification \citep{2025wny_classification_johansson}. 

Previous work on SN~2025wny confirms it is a gravitationally lensed, H-poor SLSN-I \citep{Taubenberger25, Johansson25}; for details on the lens modeling, see \citet{Ecker26} and \citet{Storfer26}. Throughout this work, we adopt the lens model parameters of \citet{Storfer26}. These include magnifications of $\mu_A=-8.8^{+1.9}_{-1.6}$ for image A (although \citealp{Storfer26} note that this is likely underestimated; see \S\ref{sec:Lightcurves_Image_A}), $\mu_B=6.1^{+0.4}_{-0.4}$ for image B, $\mu_C=-3.1^{+0.2}_{-0.2}$ for image C, and $\mu_D=3.3^{+0.1}_{-0.1}$ for image D. The locations of images A--D, along with the two lens galaxies, are shown in the KCWI spectral cube in Figure~\ref{fig:Cube_Image}.

We also report initial time delays from this lens model, which are sufficient to analyze the physics of SN~2025wny but likely too imprecise for a detailed cosmological analysis. We follow \citet{Storfer26} by modeling the total mass of the system with a singular isothermal ellipsoid, a singular isothermal sphere, and external shear, and perform parameter estimation via  {\tt lenstronomy} \citep{birrer2018} and the Gravitational Lens Efficient Explorer (GLEE; \citealp{suyu20101, suyu20121}); see \citet{Storfer26} for additional details. The time delays are reported relative to image D, the leading image, and are $\Delta t_{AD} \equiv t_A - t_D = -65.6_{-0.5}^{+0.5}$~days for image A, $\Delta t_{BD} = -54.3_{-1.5}^{+1.0}$~days for image B, and $\Delta t_{CD} = -67.5_{-0.9}^{+0.8}$~days for image C, with image D having the latest phase. 

Initial studies from \citet{Taubenberger25} and \citet{Johansson25} reported two different redshifts, $z=2.008$ and $z=2.010$, respectively. In this work, we adopt $z=2.0155$ based on host-galaxy Ly$\alpha$, H$\alpha$ and [\ion{O}{3}] $\lambda5007$ emission features (see \S\ref{sec:spec_analysis_host}; \citealp{Li26_2025wny} also find $z = 2.015$ from a similar analysis). 

\subsection{Photometry}\label{sec:data_phot}

Panoramic Survey Telescope and Rapid Response
System (Pan-STARRS; \citealp{Chambers16}) light curves for SN~2025wny were observed as part of the Young Supernova Experiment \citep[YSE;][]{Jones21} and through a Pan-STARRS Director's Discretionary Time request (PI: Storfer). We use \texttt{photpipe} \citep{Rest05} to perform point-spread function (PSF) photometry centered on each source. Briefly, {\tt photpipe} begins with images from each epoch, reduced by the image-processing pipeline (IPP; \citealp{Magnier20}) at the University of Hawai\okina i. It then measures the image zeropoint by comparing the photometry of stars in the field to the Pan-STARRS catalog photometry \citep{Magnier20}, and performs difference imaging using templates from the 3$\pi$ survey \citep{Chambers16}. We then use the difference-image detections to determine the weighted-average centroid of each point source and perform forced photometry at these locations using {\tt DOPHOT} \citep{Schechter93}. For additional details, see \citet{Jones21, Aleo23}.

We also obtain Zwicky Transient Facility (ZTF; \citealp{Bellm19}) photometry from the ZTF forced photometry service \citep{Masci19, Masci23}\footnote{The online documentation is available at this \href{https://irsa.ipac.caltech.edu/data/ZTF/docs/ztf_forced_photometry.pdf}{website}.} at the location of image A. 

\subsection{Spectroscopy}\label{sec:data_spec}

\begin{figure}
\includegraphics[width=\linewidth]{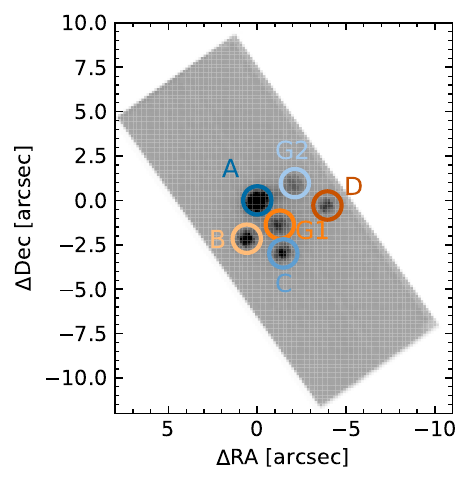}
\caption{Example image from KCWI spectral cube on 2025-11-23 ($\sim$21~rest-frame~days after Image A maximum light). Each SN~2025wny and lens galaxy image is marked by a labeled circle. We recover sources A--D but do not recover the previously reported source E \citep{Ecker26, Storfer26}.}
\label{fig:Cube_Image}
\end{figure}

\begin{deluxetable}{cccc}
\tablecaption{Log of Optical and NIR Spectroscopic Observations.\label{tab:spec-log}}
\tablewidth{0pt}
\tablehead{
\colhead{UT Date} & \colhead{MJD} & \colhead{$R$} & \colhead{Slice Width} \\
\colhead{} & \colhead{} & \colhead{[Blue/Red]} & \colhead{(arcsec)}
}
    \startdata
        \cutinhead{KCWI}
        2025-10-28 & 60976.6 & 3600/2000 & 0.35 \\
        2025-11-16 & 60995.6 & 1800/1000 & 0.70 \\
        2025-11-23 & 61002.6 & 3600/2000 & 0.70 \\
        2025-11-26 & 61005.5 & 1800/1000 & 0.35 \\
        2025-12-16 & 61025.5 & 3600/2000 & 0.35 \\
        2025-12-23 & 61032.4 & 3600/2000 & 0.35 \\
        2026-01-19 & 61059.3 & 1800/1000 & 0.70 \\
        \cutinhead{NIRES}
        2025-12-13 & 61022.4 & 2700 & \nodata
    \enddata
\tablecomments{The time of maximum light for image A is MJD 60940.4.}
\end{deluxetable}

We used the Keck Cosmic Web Imager (KCWI; \citealp{Morrissey18, McGurk24}) to observe SN~2025wny on seven separate nights. Because KCWI is an integral-field spectrograph, we obtain four spectra of SN~2025wny per pointing at each epoch, yielding a total of 28 spectra. Table~\ref{tab:spec-log} presents a log of KCWI spectroscopic observations. Figure~\ref{fig:Cube_Image} shows an example white-light image from one of the cubes, showing the 4 lensed sources. This makes SN~2025wny the best-sampled SLSN-I in the rest-frame UV to date, with wavelength coverage reaching as blue as Ly$\alpha$ and temporal coverage extending over two months after maximum light, more than twice that of other SLSNe-I \citep[e.g., SN~2011ke;][]{Quimby2018}. 

We used the reduced data cubes available via the Keck Observatory Archive from the KCWI Data Reduction Pipeline \citep[DRP;][]{Neill2023_KCWIDRP}, but we did not use the pipeline's sky subtraction and flux calibration, which are known to be suboptimal. Instead, we performed manual sky subtraction and flux calibration using the differential atmospheric refraction-corrected cubes. To do so, we extracted 1D spectra from each lensed image, each lensing galaxy source, and the background sky from each data cube. Next, we median-combined each source, and then subtracted the final median-combined sky spectrum from each final source spectrum. Flux calibration on the sky-subtracted frames used CALSPEC reference stars observed during the same night. 

We also obtained a near-infrared (NIR) spectrum of image A of SN~2025wny using the NearInfraRed Echellette Spectrometer (NIRES; \citealp{Wilson04_NIRES}) on the 10.0m Keck-II telescope. The NIRES spectrum was obtained on MJD 61022.4, corresponding to $\sim$$+24$~days after maximum light for image A. We reduced the NIRES data using a modified version of \texttt{spextool} \citep{Cushing04}. We used an A0V star to perform telluric corrections with the \texttt{xtellcor} package \citep{Vacca03}. 

\section{Photometric Analysis}\label{sec:phot_analysis}
Given the struggles of lens models in predicting the magnification of image A, we compute its magnification directly from our light curves We present our technique and results in \S\ref{sec:Lightcurves_Image_A}, and use the corrected magnification estimate in subsequent analysis. In \S\ref{sec:lc_params}, we estimate the time of maximum light in the image A rest frame and compare the absolute magnitude of SN~2025wny with those of lower-redshift SLSNe. Finally, \S\ref{sec:MOSFiT} presents magnetar modeling of the light curve and interprets the results.

\subsection{Image A Magnification Estimation}\label{sec:Lightcurves_Image_A}

\begin{figure}
\includegraphics[width=\linewidth]{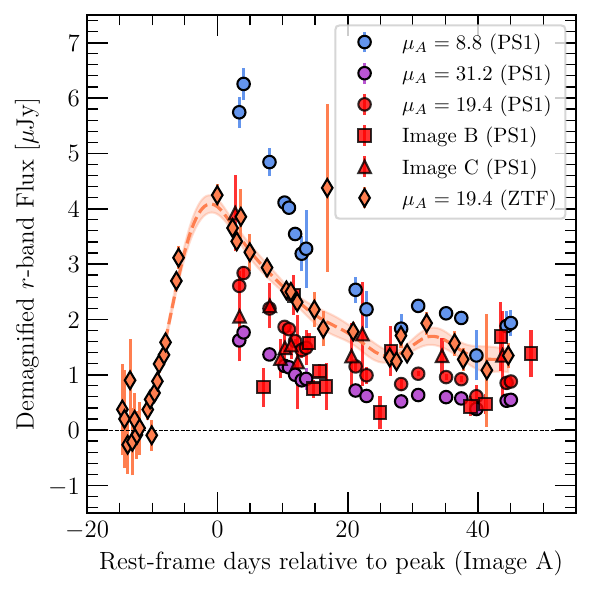}
\caption{Demagnified, rest-frame Pan-STARRS $r$-band light curves of SN~2025wny image A encompassing the lensing magnification range from \citet{Storfer26}, $8.8 < |\mu_A| < 31.2$ (blue and purple, respectively), alongside the best-fit value derived here (red circles; $|\mu_A|=19.4$).  The demagnified ZTF $r$-band light curve is shown for comparison, along with the GP fit used to measure time of maximum light (blending with images B--D likely explains the increased flux relative to Pan-STARRS; \citealp{Li26_2025wny}). The demagnified image B (red squares; $|\mu_B| = 6.1$) and image C (red triangles; $|\mu_C| = 3.1$) fluxes are used to determine the best-fit image A magnification.}
\label{fig:Phot_demagexample}
\end{figure}

\citet{Storfer26} report that while the predicted magnifications of images B--D are broadly consistent with their photometry, image A is anomalously a factor of 2--3 brighter than what their lens modeling predicts, implying a true magnification between $\sim$17--26 rather than $\sim$9. For this reason, we fit for the flux ratio between the time-delay-corrected, rest-frame light curves of images A and C to determine the magnification ratio and thus obtain an empirical magnification estimate for image A. Although image B is brighter than image C, we use image C because it aligns better temporally with image A, providing a better constrained fit. Calibrating image A against image C helps remove time- and color-constant differences in dust extinction or microlensing, since we tie A to C. The fitting is performed as follows. 

First, we compare each image C epoch directly to the corresponding image A epoch, or to a weighted average of all image A epochs within one rest-frame day. This is done in the Pan-STARRS $g$-, $r$-, $i$- and $z$-bands, which cover 0--50~days after $r$-band maximum light. We exclude the $y$-band because of sparse data coverage. This yields 8, 16, 18, and 10 matched image A and C pairs in the $g$, $r$, $i$, and $z$ bands, respectively. We model the image C flux as a multiple of the image A flux, i.e., $F_c = s\times F_A$, where $s \equiv \mu_C/\mu_A \equiv \frac{1}{R}$, with $R$ being the magnification ratio. We fit all parameters simultaneously, with $s$ held constant across all bands, and estimate uncertainties from 10,000 Monte Carlo realizations that resample the photometric noise in the image A and C light curves. 

The joint fit gives $R = 6.3^{+0.3}_{-0.3}$.  We include a parameter to scale the flux uncertainties, $f_{\rm int}$,  and find that the errors are underestimated by an average of $f_{\rm int} = 23.6\%$. Our fit gives $\chi^2/{\rm dof} = 1.05$ for the 52 matched epochs. Fitting each filter independently returns $R = 4.6^{+1.3}_{-0.9}$, $7.0^{+0.7}_{-0.6}$, $6.1^{+0.5}_{-0.4}$, and $7.1^{+0.6}_{-0.5}$ in $g$, $r$, $i$, and $z$. These values agree across bands within their uncertainties at the $\sim$1.8$\sigma$ level.

We convert the fitted ratio to an absolute magnification using the \citet{Storfer26} lens model $\mu_C$, drawing $\mu_C$ from $\mathcal{N}(3.1, 0.2)$ within each Monte Carlo realization to include the lens-model uncertainty. Our final fit yields $\mu_A = 19.4^{+1.6}_{-1.6}$, in agreement with the range implied by the magnification anomaly reported by \citet[which used NIRC2 photometry]{Storfer26} and the estimate from \citet[their \S3.1]{Li26_2025wny}. We adopt this value as the magnification estimate for image A. 

To check the result, we repeat this exercise for image B and find the fit is fully consistent but more poorly constrained ($\mu_A = 20.5^{+2.9}_{-2.7}$), likely because of less overlap between image A and image B.
%We note that this agreement does not independently confirm the ``overmagnification'' of image A. 
However, we caution that all the magnifications are correlated because they are drawn from the same lens model. Further lens modeling for this system using additional data may shed more light on the true magnification of image A. 

We note that these results could be affected by chromatic microlensing or substructure within the lensing system.  The possibility of chromatic microlensing, which could bias modeling constraints by changing the color of the SN, was examined by \citet{Townsend26_2025wny}.  They varied their microlensing parameters independently between bands and found that doing so lowers the Bayesian evidence ($\Delta \log Z = -3$) and does not improve the fit.  Given that our single-band magnification estimates are also consistent with each other within the uncertainties, we do not consider chromatic microlensing further in this work, though we note that it may subtly affect the derived parameters in Section \ref{sec:MOSFiT}.

On the other hand, time-dependent microlensing could also affect these estimates.  \citet{Townsend26_2025wny} model the magnification difference over time using a sigmoid function, finding a microlensing onset time that corresponds approximately to the first epoch of Pan-STARRS data ($t_{0,A} = 60953^{+17}_{-18}$~days, while the first Pan-STARRS data point is at MJD = 60951.6).  Our magnification estimates are therefore unlikely to be highly affected by the time-dependent microlensing found by \citet{Townsend26_2025wny}.  As an additional check, we calibrate image A against image C separately over rest-frame phases 0–20 and 20–50 d, finding $\mu_A = 22.5 \pm 2.1$ and $15.4^{+2.2}_{-2.0}$, respectively (adopting $\mu_C = 3.1 \pm 0.2$ from \citealp{Storfer26}; the $\mu_C$ uncertainty is common to both). We find a corresponding change in A/C of $\Delta m = 0.41 \pm 0.13$~mag, significant at the 3$\sigma$ level.  Because our later-time data are more sparse and lower-S/N, and this change in magnification lies within our overall uncertainties, we adopt the global fit in this work but note that this is additional evidence for time-dependent microlensing beyond what was found in \citet{Townsend26_2025wny}.  We also restrict our primary modeling analysis below to phases $<+20$~days, which limits our sensitivity to time-dependence in the magnification.

Figure~\ref{fig:Phot_demagexample} shows the YSE $r$-band light curves for images A (with varying magnifications), B, and C along with the ZTF $r$-band light curve. Our flux values for image A (with the corrected magnification) and images B/C agree reasonably well. We note that the ZTF image A light curve likely includes additional flux contributions from other images of SN~2025wny, as the ZTF PSF size is comparable to the separation between images \citep{Li26_2025wny}.

\subsection{Light Curve Parameters}\label{sec:lc_params}

We fit the ZTF light curve with a Gaussian Process (GP) to estimate the time of maximum light in image A, shown in Figure~\ref{fig:Phot_demagexample}. We omit data with $S/N < 3$ from the fit and use a Mat\'ern $\nu=5/2$ kernel, with the flux uncertainties as the variance. We adopt a 10-day length scale in the rest frame, corresponding to the observed rise time (which may differ significantly from the true physical rise time). We find the maximum $r$-band magnitude occurred on MJD $60940.4\pm0.2$. The demagnified peak flux is $\sim$4~$\mu\mathrm{Jy}$, or $\sim$22.4~mag (AB system), corresponding to an observed absolute AB magnitude of $M^{\mathrm{obs}}_r\approx-23.6~\mathrm{mag}$.  

At $z\sim2$, the observed $r$-band samples approximately  1800--2300~\AA\ in the rest-frame, overlapping the \emph{Swift} UVOT $UVW2$ and $UVM2$ bands. To convert the observed magnitude to a rest-frame $r$-band magnitude, we combine the $+21.6$~day optical spectrum (chosen for its high S/N) and NIR spectrum into one SED. Fortunately, the $g-r$ color of the 2025wny SED changes by only $\sim$0.08~mag between maximum light and $+$21.6 days, so this $K$ correction is a reasonable approximation at maximum light. We perform synthetic photometry on the observed and rest-frame $r$-band regions of the SED and compute the $K$ correction following \citet[their Equation 12]{Hogg02}.
From this, we derive a $K$ correction of $-1.22$~mag.\footnote{We note that the SLSNe-I $K$-corrections from \citet{gomez2024} are not well behaved at such blue wavelengths, and would imply an unphysically high luminosity (see their \S4.5).}

Using this $K$-correction, the rest-frame $r$-band absolute magnitude is $M^{\mathrm{rest}}_r\approx-22.4~\mathrm{mag}$, with an uncertainty dominated by lens modeling of about 0.5~mag. \citet{gomez2024} report similarly luminous rest-frame $r$-band magnitudes for the high-$z$ SLSNe DES16C2nm at $z=1.998$ ($M^{\mathrm{rest}}_r=-22.67\pm0.53~\mathrm{mag}$; \citealp{smith2018}) and SCP06F6 at $z = 1.189$ ($M^{\mathrm{rest}}_r=-22.55\pm0.21~\mathrm{mag}$; \citealp{quimby2011}). At lower redshift, SN~2020dbl ($z = 0.398$; \citealp{Chen23_I}) has a rest-frame absolute magnitude of $M^{\mathrm{rest}}_r=-22.44\pm0.05~\mathrm{mag}$ \citep{gomez2024} and SN~2002gh ($z = 0.365$) has $M_V = -22.40\pm0.02~\mathrm{mag}$ \citep{Cartier22}. Even given the large lensing magnification uncertainty, the peak absolute magnitude of SN~2025wny places it at the high-luminosity end of the SLSN-I population.  For additional confirmation of this high (albeit uncertain) luminosity, see the 2025wny light curve modeling using MOSFiT in Section \ref{sec:MOSFiT} below; MOSFiT applies its own $K$ corrections to our data.

\begin{figure}
\includegraphics[width=\linewidth]{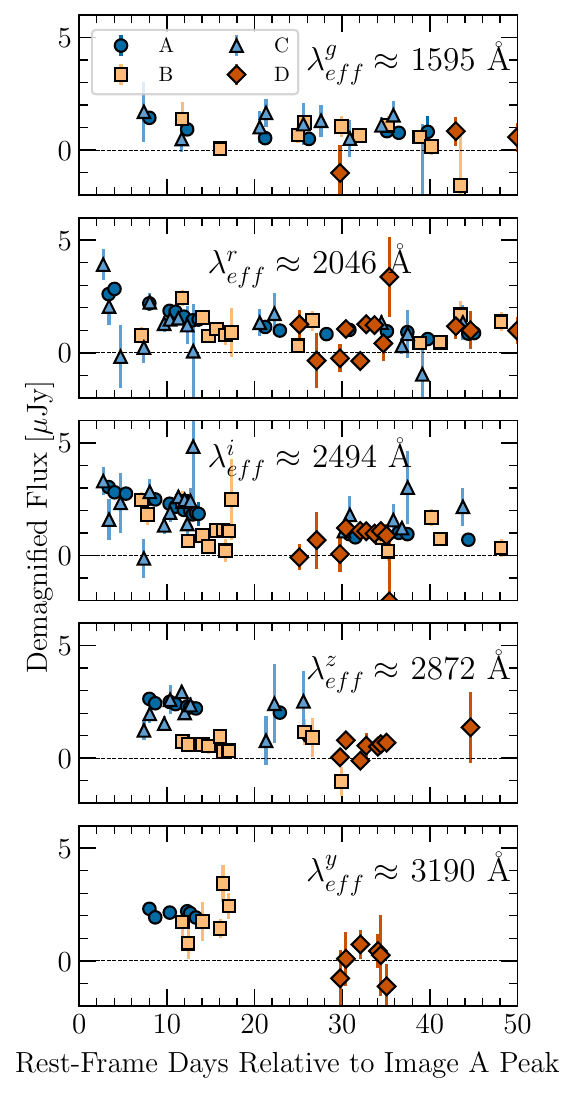}
\caption{The rest-frame, demagnified YSE light curves of SN~2025wny images A--D, separated by filter. The rest-frame effective wavelengths for each observed bandpass are shown in the upper right of each panel. }
\label{fig:Phot_lightcurve}
\end{figure}

Figure~\ref{fig:Phot_lightcurve} shows the demagnified Pan-STARRS light curves for SN~2025wny, separated by image. The data show a plateau in the $gri$ bands, and possibly a slight rise in the $gr$ bands (see Figure \ref{fig:mosfit_lc}; 20--40 days after maximum light). Both ZTF and Pan-STARRS detect the plateau epochs with reasonable S/N over multiple epochs, and visual inspection of the Pan-STARRS images confirms the detections.  Such post-peak plateaus or rises in SLSNe-I are not uncommon --- \citet{Hosseinzadeh22_SLSN} estimate that 44--76\% of SLSNe cannot be explained solely by a smooth magnetar spin-down (see also \citealp{Chen23_II, gomez2024}) ---  and may arise from CSM interaction \citep[e.g.,][]{Inserra2017, Li20_SLSN_CSM, Chen23_II} or additional energy injection from the central engine from fallback accretion \citep[e.g.,][]{Dexter13, Metzger2014, Metzger2018}. Alternatively, \citet{Farah26} suggest some undulations could arise from Lense-Thirring precession of an accretion disk around the newly formed magnetar. Of the four SLSNe-I that \citet{Farah26} claim show Lense-Thirring precession, the light curve of SN~2025wny most closely resembles that of SN~2018kyt, while the others differ from SN~2025wny in their rise and peak light-curve shapes. Our data do not allow a conclusive determination of the plateau source, and \citet{Li26_2025wny} likewise cannot definitively determine the cause of the plateau, with their observations supporting both CSM-interaction and central-engine energy-injection scenarios.

\subsection{Light Curve Modeling}\label{sec:MOSFiT}

\begin{figure}
    \centering
    \includegraphics[width=\linewidth]{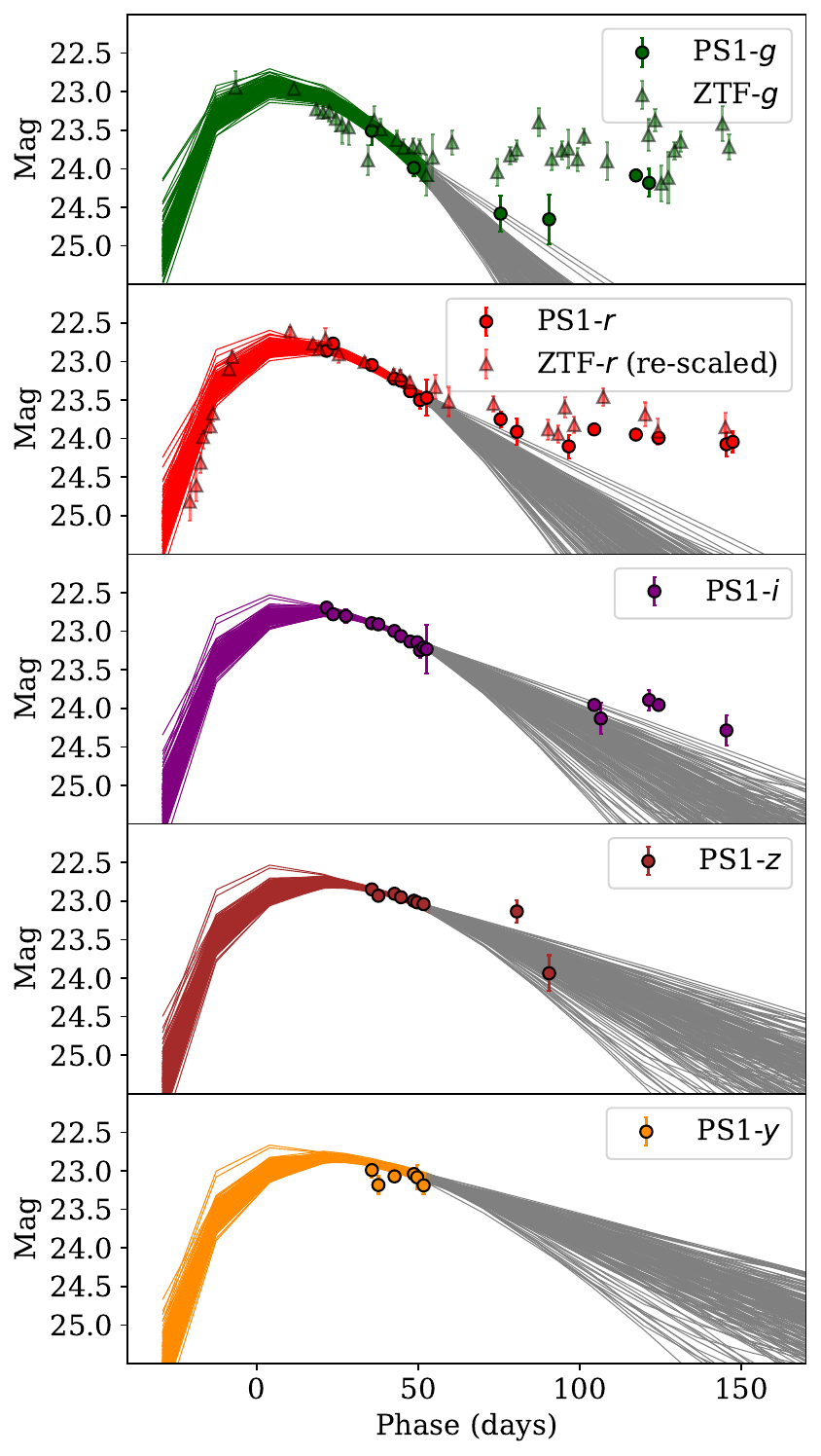}
    \caption{The 2025wny light curve from Pan-STARRS (circles), with light-curve realizations from MOSFiT posterior sampling as solid lines.  ZTF data (triangles) are shown for comparison but are not used in the fit; the ZTF $r$ band has also been re-scaled to match the Pan-STARRS data \citep{Li26_2025wny}.  Colored lines show MOSFiT models in regions where the data are fit, while gray lines show the extrapolation of the near-peak model.}
    \label{fig:mosfit_lc}
\end{figure}

We model the physical parameters of SN~2025wny with the Modular Open-Source Fitter for Transients (MOSFiT; \citealp{Guillochon18_MOSFiT}).  MOSFiT uses the {\tt emcee} MCMC ensemble sampler to fit the data, and we use $20000$ MCMC steps to ensure reliable posterior sampling. We assume a magnetar central engine as the dominant power source, with additional contributions from radioactive $^{56}$Ni decay (the ``{\tt slsnni}" model).  Magnetar models have been shown to reproduce SLSN light curves and spectra reliably \citep[e.g.,][]{Kasen10b, mazzali2016, Nicholl17, Hinkle25_SwiftFixII}, with a relatively minimal parameter set compared to alternatives such as CSM-interaction models.

Because the image A--D fluxes are blended within the ZTF photometry, we use only the Pan-STARRS data to estimate the {\tt slsnni} model parameters. However, because the Pan-STARRS data begin after maximum light, we use the ZTF data to place a prior on the explosion time. We first rescale the blended ZTF $r$-band photometry to match Pan-STARRS, finding a scaling consistent with the magnitude offset reported in \citet[their Figure 2]{Li26_2025wny}. Finally, we fit the {\tt slsnni} model to the combined ZTF and Pan-STARRS data, and use the resulting constraints on $t_{exp}$ to set the bounds for the subsequent Pan-STARRS-only fit.

We find that MOSFiT can produce reasonable fits to the near-peak data for SN~2025wny, but is unable to model the later-time plateau (which may even include a slight rise); if we attempt to fit all data simultaneously, the model is flatter than the light curve near peak but still declines more steeply than the data at later times.  While the MOSFiT {\tt slsnni} model can reproduce a slower decline rate at late times via the $\gamma$-ray opacity parameter $\kappa_{\gamma}$, it is unable to match the late-time data, and therefore it is likely that another energy source is needed to explain the late-time behavior. For this reason, we fit only data prior to the observed plateau, prior to ${\rm MJD}=61000$ (or within 20~rest-frame~days after image A $r$-band maximum).

Figure \ref{fig:mosfit_lc} shows the \texttt{MOSFiT} realizations from our fitting, with the predicted model at late times (gray) showing strong discrepancies with the late-time data (particularly in the $g$ band). For SN~2025nwy, \citet{Li26_2025wny} favor fallback accretion onto the central engine. We also note that these additional power sources could contribute to the near-peak flux.  While our data are too sparse to distinguish between these models, we proceed with a magnetar central engine model because it allows useful comparisons to the literature sample; our primary goal in this subsection is to assess whether a magnetar model can reasonably explain the near-peak emission for SN~2025wny.

\begin{deluxetable*}{lccrl}
\tabletypesize{\footnotesize}
\tablecaption{\texttt{MOSFiT} Parameters for the SLSN$+$Ni Model\label{tab:mosfit}}
\tablewidth{0pt}
\tablehead{ \colhead{Parameter} & \colhead{Units} & \colhead{Prior Range} & \colhead{Parameter Value} & \colhead{Definition} }
    \startdata
        $M_{\rm ej}$&$M_{\odot}$&$0.1,100$&$4.267^{+1.836}_{-1.281}$&Ejecta mass\\
        $f_{\rm Ni}$&&$\log(10^{-5},1.0)$&$0.010^{+0.548}_{-0.010}$&Nickel mass as a fraction of ejecta mass\\
        $V_{\rm ej}$&${\rm km~s^{-1}}$&$10^3,10^5$&$6703^{+1335}_{-996}$&Ejecta velocity\\
        $M_{\rm NS}$&$M_{\odot}$&$1.0,3.0$&$2.473^{+0.370}_{-0.577}$&Neutron star mass\\
        $P_{\rm spin}$&ms&$0.7,30$&$4.104^{+0.711}_{-1.008}$&Magnetar spin\\
        $B_{\perp}$&$10^{14}~{\rm G}$&$\log(10^{-3},15.0)$&$2.347^{+1.045}_{-0.818}$&Magnetar magnetic field strength\\
        $\theta_{\rm PB}$&rad&$0,\pi/2$&$1.081^{+0.378}_{-0.397}$&Angle of the dipole moment\\
        $t_{\rm exp}$&MJD&$60891,60894$&$60892.8^{+0.9}_{-1.1}$&Explosion time\\
        $T_{\rm min}$&K&$3000,10000$&$6881^{+2456}_{-2896}$&Photosphere temperature floor\\
        $\lambda_0$&\AA&$2000,6000$&$2838^{+74}_{-73}$&Wavelength below which flux suppression begins\\
        $\alpha$&&$0,5$&$0.744^{+0.168}_{-0.249}$&Slope of the flux suppression\\
        $n_{H,{\rm host}}$&${\rm cm^{-2}}$&$\log(10^{16},10^{23})$&$17.589^{+1.524}_{-1.236}$&H\,I column density in the host galaxy\\
        $\kappa$&${\rm cm^2~g^{-1}}$&$0.01,0.34$&$0.019^{+0.016}_{-0.007}$&Optical opacity\\
        $\kappa_{\gamma}$&${\rm cm^2~g^{-1}}$&$\log(10^{-2},10^1)$&$0.229^{+3.718}_{-0.208}$&$\gamma$-ray opacity\\
        $\sigma$&&$\log(10^{-3},10^2)$&$0.005^{+0.013}_{-0.003}$&Uncertainty scale giving $\chi^2/\nu = 1$\\
    \enddata
\tablecomments{Prior ranges noted in log have a log-flat prior, while other priors have a flat prior. The prior range on $t_{exp}$ is determined using a MOSFiT fit to ZTF$+$Pan-STARRS data; aside from this prior, the reported parameters use Pan-STARRS data only.  The parameter descriptions are adapted from \citet{gomez2024}. Reported \texttt{MOSFiT} parameters are the median value and 1$\sigma$ uncertainties (i.e., the $16^{\rm th}$ and $84^{\rm th}$ percentiles).}
\end{deluxetable*}

Table~\ref{tab:mosfit} lists the full set of model parameters, including our best-fit measurements, and the Appendix includes a corner plot of the posterior distributions.  Our data cannot significantly constrain energy input from radioactive decay or $\gamma$-ray opacity, with the Nickel mass fraction ($f_{\rm Ni}$) and $\kappa_{\gamma}$ remaining largely unconstrained within the priors (given these poor constraints, in our final parameter estimation we use an upper limit of $10~{\rm cm^2~g^{-1}}$ on $\kappa_{\gamma}$ that is somewhat lower than several other SLSN studies; see \citealp{gomez2024} for additional justification for lower $\kappa_{\gamma}$ values).  The neutron-star mass, $M_{\rm NS}$, is also poorly constrained at $M_{NS} = 2.5^{+0.4}_{-0.6}~M_{\odot}$, while the magnetar spin ($P_{\rm spin}$) and the optical opacity ($\kappa$) favor values near the bottom of their respective prior ranges.  We obtain good constraints within the prior bounds on most other parameters, including flux suppression parameters $\alpha$ and $\lambda_0$,  
an ejecta mass of $4.3^{+1.8}_{-1.3}~M_{\odot}$, 
a magnetic field strength of $2.3^{+1.0}_{-0.8}\times10^{14}~{\rm G}$, 
a spin period $4.1^{+0.7}_{-1.0}~{\rm ms}$, 
and an ejecta velocity of $6.7^{+1.3}_{-1.0}\times10^3~{\rm km~s^{-1}}$. While the 50th percentile of the neutron star mass measurement is higher than that of every other SLSN-I in \citet{Hinkle25_SwiftFixII} and \citet{Nicholl17}, it is consistent with these studies within 1$\sigma$ ($\sim$1.9~$\mathrm{M}_\odot$). The extreme upper end of the mass posterior exceeds the maximum non-spinning neutron star mass \citep[which depends heavily on the neutron star equation of state, e.g.,][]{Ye22_NSmass}. 

We note that the $r$-band rise in the blended ZTF photometry is slightly discrepant with our baseline Pan-STARRS-only fit (see the second-from-top panel in Figure \ref{fig:mosfit_lc}). However, when we include the blended ZTF photometry in the fit, we can reproduce the rise exactly.  Because time-dependent blending could be an issue in these data, we do not choose to use these blended data to produce our baseline MOSFiT results; however, we note that these combined fits yield consistent parameters.  The constraints on $M_{NS}$ and opacity parameters still largely recover the prior when blended ZTF data are used, while the ejecta mass, velocity, and magnetic field strength are all consistent within $\sim$1$\sigma$.

MOSFiT also models UV line blanketing in SLSNe with a linear flux-suppression term, $\alpha$, which applies at wavelengths bluer than the $\lambda_0$ parameter. Our results suggest flux suppression beginning below $2838^{+74}_{-73}$~\AA, with a slope of $\alpha = 0.74^{+0.17}_{-0.25}$.  \citet{gomez2024} find population averages of $\lambda_0 = 3400^{+100}_{-700}$ and $\alpha = 1.7^{+1.5}_{-1.3}$, where error bars here give the parameter ranges encompassing 68\% of the population.  The flux suppression in SN~2025wny starts $\sim$1$\sigma$ bluer, and flux suppression blueward of that wavelength falls in the bottom 16\% of the population.  This is consistent with the lack of line blanketing seen by both \citet{Li26_2025wny} and our spectral analysis in Section \ref{sec:spec_analysis}.

\begin{figure*}
\includegraphics[width=\textwidth]{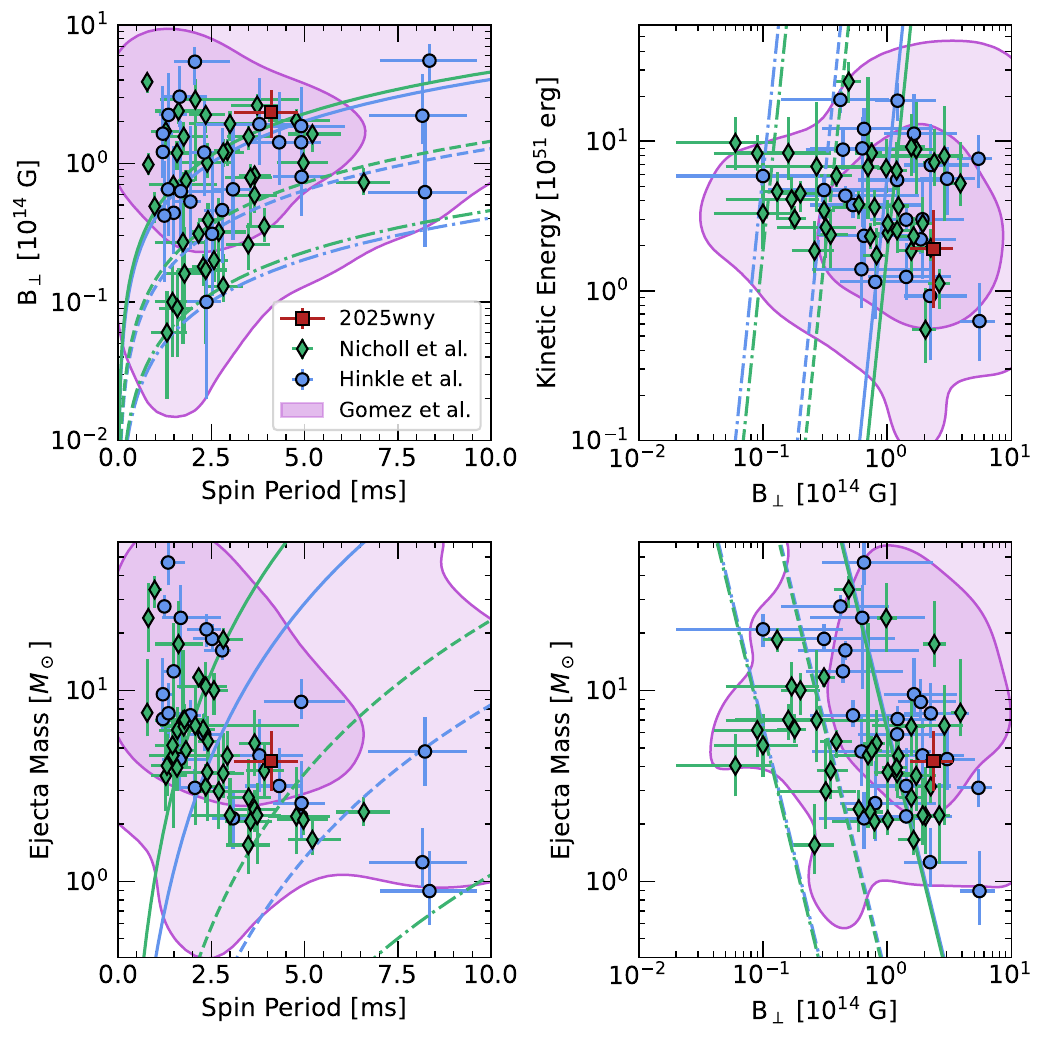}
\caption{Comparison of \texttt{MOSFiT} parameters for SN~2025wny (red square) compared against those from \citet[blue circles]{Hinkle25_SwiftFixII}, \citet[green diamonds]{Nicholl17}, and $1\sigma$ and $2\sigma$ contours from \citet[purple]{gomez2024}. We also show lines of constant ratio (10, 1, and 0.1 as dash-dot, dashed, and solid lines, respectively) between the magnetar spin-down and diffusion timescales.}
\label{fig:Phot_mosfit_comp}
\end{figure*}

Figure~\ref{fig:Phot_mosfit_comp} shows the \texttt{MOSFiT} parameters compared to SLSNe samples from \citet{Nicholl17} and \citet{Hinkle25_SwiftFixII}. SN~2025wny has magnetar model parameters consistent with the lower-redshift SLSNe sample, with a spin period slightly longer than the sample average. The ejecta mass and kinetic energy are near the low end of the observed population, but their uncertainties place them within $\sim$$1\sigma$ of the population average. \texttt{MOSFiT} also returns a bolometric peak luminosity of $L_\mathrm{peak} = (8.1\pm2.5)\times10^{44}~\mathrm{erg~s^{-1}}$ and a total rest-frame, radiated energy of $E_\mathrm{rad} = 1.9^{+0.7}_{-0.5}\times10^{51}$~erg. This value is statistically consistent with the upper $\sim$10\% of the $E_\mathrm{rad}$ range (see their Figure 4). In particular, the highest-redshift SLSNe-I in the \citet{gomez2024} sample are more luminous (likely due to Malmquist bias), and SN~2025wny is consistent with that population. \citet{Li26_2025wny} estimate a peak luminosity of $L_{\mathrm{peak}} = 4.4^{+1.2}_{-0.8}\times10^{44}$~erg~s$^{-1}$, consistent with our measurement at the 1.3$\sigma$ level (despite their lower magnification estimate for Image A).

\section{Spectroscopic Analysis}\label{sec:spec_analysis}

\subsection{Spectral Time Series and Feature Identification}\label{sec:phot_analysis_timeseries}

\begin{figure*}
\includegraphics[width=\textwidth]{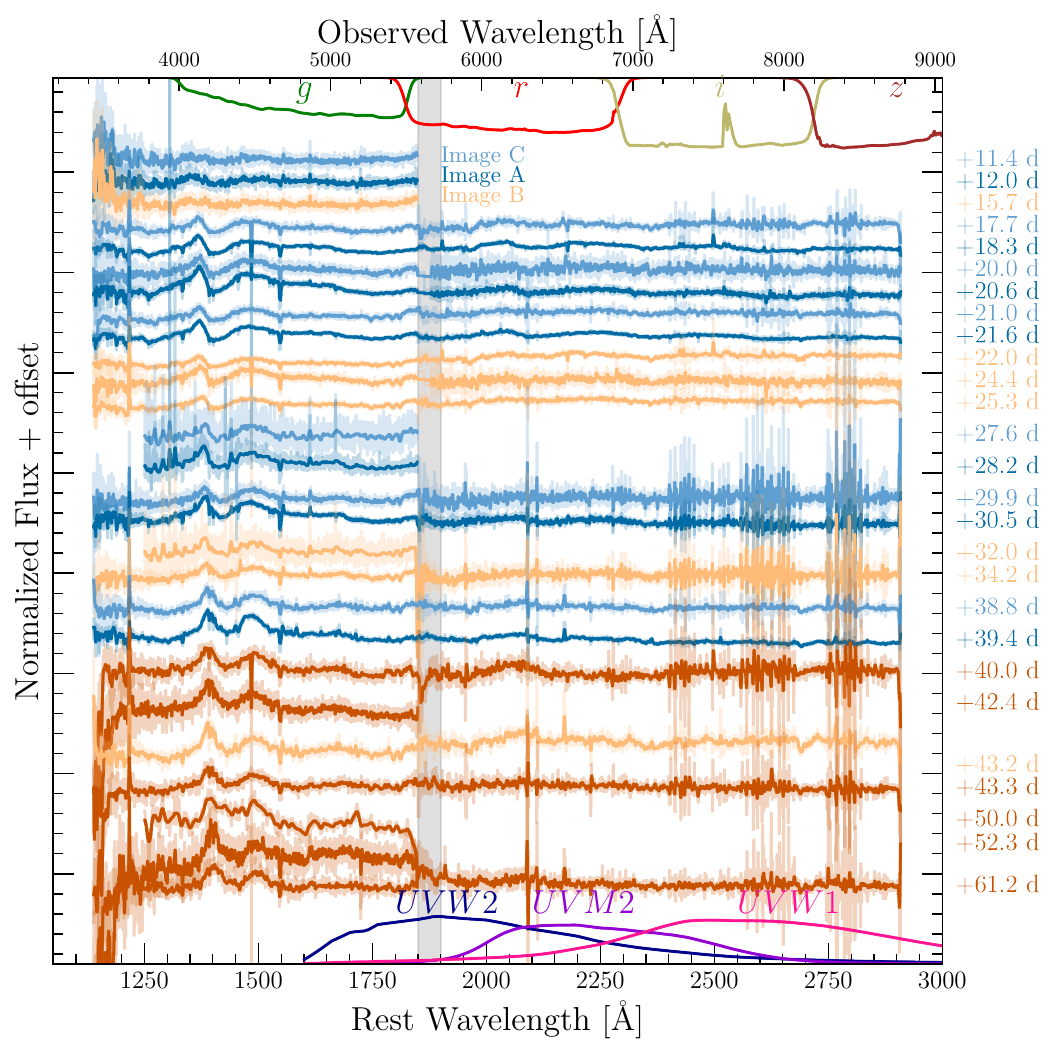}
\caption{The rest-frame, time-dilation-corrected spectral time series for SN~2025wny. Low-S/N regions are masked, and the grey shading denotes the overlap between the blue and red KCWI sides. The observed Pan-STARRS bandpasses are shown at the top of the figure, and the rest-frame \emph{Swift} UVOT bandpasses are shown at the bottom. The more opaque lines are smoothed spectra. We highlight that even the low-opacity ``un-smoothed'' spectrum, the image D spectrum at $50.0$~d, has been smoothed due to the low $S/N$ for that epoch.}
\label{fig:Spectra_timeseries}
\end{figure*}

Figure \ref{fig:Spectra_timeseries} presents the spectral time series of SN~2025wny from our KCWI observations, which covers $\sim$50 days of rest-frame evolution. Two properties of our spectral coverage are worth noting explicitly. First, while DES16C2nm is at a similar redshift and thus has similar rest-frame UV coverage to SN~2025wny \citep{smith2018}, the strong lensing magnification significantly improves our $S/N$. Second, prior to SN~2025wny, the latest rest-frame UV spectrum of an SLSN-I was at $+25$~d after maximum light (DES16C2nm, again). Thus, SN~2025wny provides the latest rest-frame UV spectra of an SLSN-I to date and the highest-$S/N$ spectra covering blueward of $\sim$1500~\AA.  This study in particular approximately doubles the available spectra of 2025wny, with a similar number of spectral epochs to \citet{Li26_2025wny}, and includes a late-time epoch $\sim$1 week later in the rest frame than was previously available.

\begin{figure*}
\includegraphics[width=\textwidth]{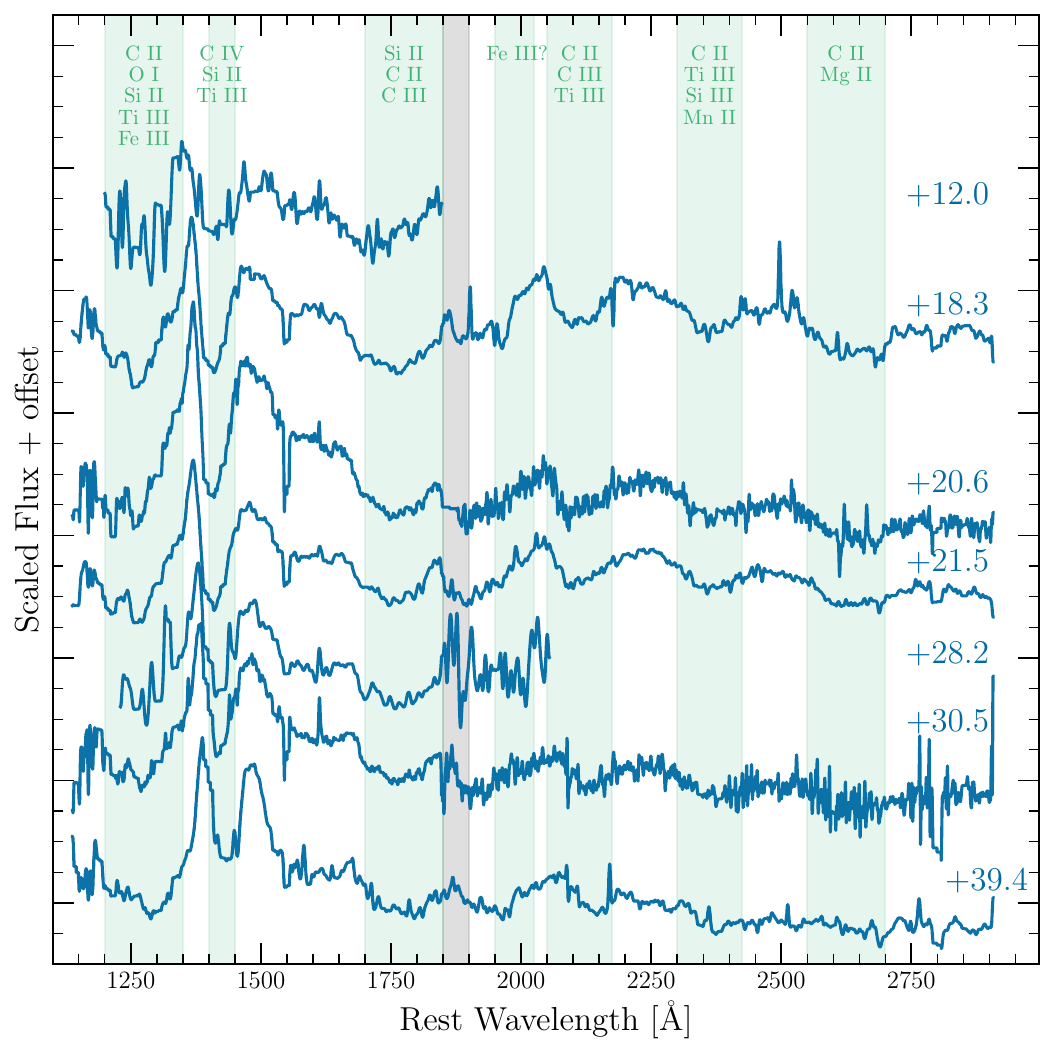}
\caption{Spectral time series of image A, the brightest and highest-S/N source of our spectroscopy. Phases are adjusted for time dilation, and spectrum colors match the scheme adopted in Figure \ref{fig:Cube_Image}. The grey shading denotes the overlap between the blue and red KCWI sides, and narrow ISM absorption lines are masked.}
\label{fig:Spectra_timeseries_source_A}
\end{figure*}

We show line identifications for the spectra of image A in Figure \ref{fig:Spectra_timeseries_source_A}; we show only image A because it has the highest signal-to-noise ratio among our spectra. We see previously reported absorption features from \citet{Taubenberger25}, \citet{Johansson25}, and \citet{Li26_2025wny}. These include, from bluest to reddest, absorption features at 
$\sim$$\lambda\lambda$1350--1400,
$\sim$$\lambda\lambda$1450--1500,
$\sim$$\lambda\lambda$1750--1800,
$\sim$$\lambda\lambda$1800--2000, 
$\sim$$\lambda\lambda$2100--2250, 
$\sim$$\lambda\lambda$2400--2500, and
$\sim$$\lambda\lambda$2600--2750.

The spectral feature identification is generally consistent between initial studies by \citet{Taubenberger25} and \citet{Johansson25}. The subsequent study by \citealt{Li26_2025wny}, whose line identifications follow \citealt{Johansson25}, and our own work also agree well. Generally, this agreement holds for the redder wavelengths, but feature identification diverges at the bluest wavelengths, likely because prior radiative transfer modeling has not been used to understand spectral line formation from $\sim$1200~\AA\ to $\sim$1800~\AA. For example, \citet{mazzali2016} focus on spectral features redder than $\sim$2000~\AA. The bluest feature they identify is an \ion{Al}{3} and \ion{Si}{3} blend at $\sim$$\lambda1700$ in SNLS-06D4eu \citep{howell2013}, but that feature was identified by \citet{Johansson25} as \ion{Si}{2}, \ion{C}{2}, and \ion{C}{3} instead.

\begin{figure}
\includegraphics[width=\linewidth]{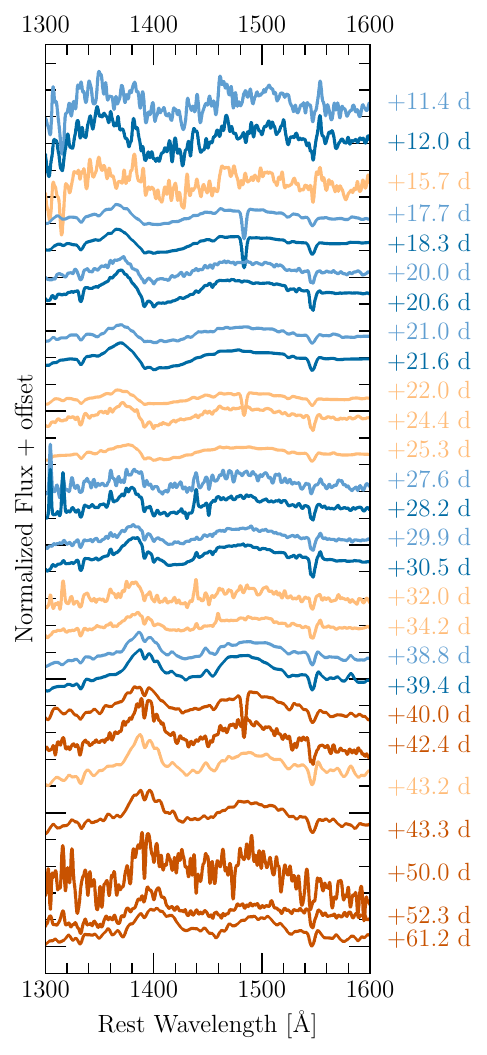}
\caption{Evolution of the line complex between 1400 and 1500 $\mathrm{\AA}$. The spectra are smoothed to improve visual clarity.} 
\label{fig:Spectra_1450evol}
\end{figure}

Figure \ref{fig:Spectra_1450evol} shows the evolution of the main post-peak FUV spectral feature between 1400--1500~\AA. \citet{Johansson25} attribute this feature in their spectrum taken $\sim$10~days after maximum light to a blend of \ion{C}{2}, \ion{O}{1}, \ion{Si}{2}, \ion{Ti}{3}, and \ion{Fe}{3}. 

\subsection{Comparison to other Type I Superluminous Supernovae}

Rest-frame UV spectroscopy of SLSN-I from the literature originates from two channels. First, direct observer- and rest-frame UV spectroscopy from \emph{HST} of nearby SLSNe ($z \lesssim0.15$), or second, ground-based optical/NIR observations of high-redshift SLSNe ($z\gtrsim0.5$). 

No other SLSN has the same level of UV spectral sampling as SN~2025wny. Roughly half of the spectra presented in this work were obtained after $+25$~days, making them incomparable to existing rest-frame UV data because of the large phase differences. Nonetheless, a small subset of spectra overlaps in phase and wavelength with our observations. This sample includes SN~2011ke \citep{Quimby2018}, PTF12dam \citep{Quimby2018}, Gaia16apd \citep{yan2017}, and DES16C2nm \citep{smith2018}.  

Generally, SLSNe-I ejecta have low iron-group element abundances \citep[e.g.,][]{mazzali2016}, resulting in fewer absorption features from these species and less line blanketing \citep[e.g.,][]{yan2017, Yan18}. SN~2025wny has conspicuously weaker/absent \ion{Fe}{3} $\sim$$\lambda\lambda1800$--2000 than SNLS-06D4eu and Gaia16apd at early times, and our spectral time series does not show the prominent absorption feature in that wavelength range. Indeed, the lack of a clear absorption feature and the lack of any time evolution of the spectral profile, which we see in other features (e.g., the \ion{C}{4}, \ion{Si}{2}, and \ion{Ti}{3} blend at $\sim$$\lambda\lambda$1450--1500), supports the conclusion that \ion{Fe}{3} $\sim$$\lambda\lambda1800$--2000 is not present in SN~2025wny. This would indicate even less Fe in the outer ejecta of SN~2025wny compared to other SLSNe. 

\begin{figure*}
\includegraphics[width=\textwidth]{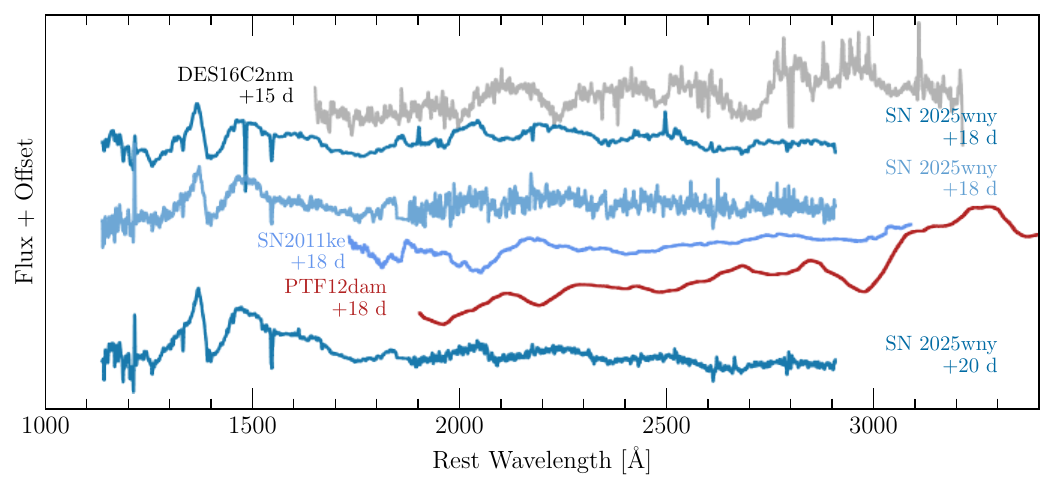}
\caption{Comparison of the rest-frame FUV spectra of H-poor SLSNe at $\sim$15--20 days after maximum light.}
\label{fig:Spectra_comp_18}
\end{figure*}

Figure~\ref{fig:Spectra_comp_18} presents a comparison between SN~2025wny, DES16C2nm, SN~2011ke, and PTF12dam at $\sim$$+18$~days after maximum light. The small sample makes comparisons tentative. Compared to the other three SLSNe-I, SN~2025wny shows similar absorption features but at higher blueshifts, suggesting either a faster ejecta velocity at this epoch (in contradiction to the rather low ejecta velocity from \texttt{MOSFiT}) or different spectral features entirely. PTF12dam has a much stronger redward slope than SN~2025wny and the other two reference SLSNE-I, which may suggest PTF12dam is more metal-rich (and thus more line-blanketed by iron-group elements). SN~2011ke in particular seems to match the almost completely flat UV slope of SN~2025wny well. 

\begin{figure*}
\includegraphics[width=\textwidth]{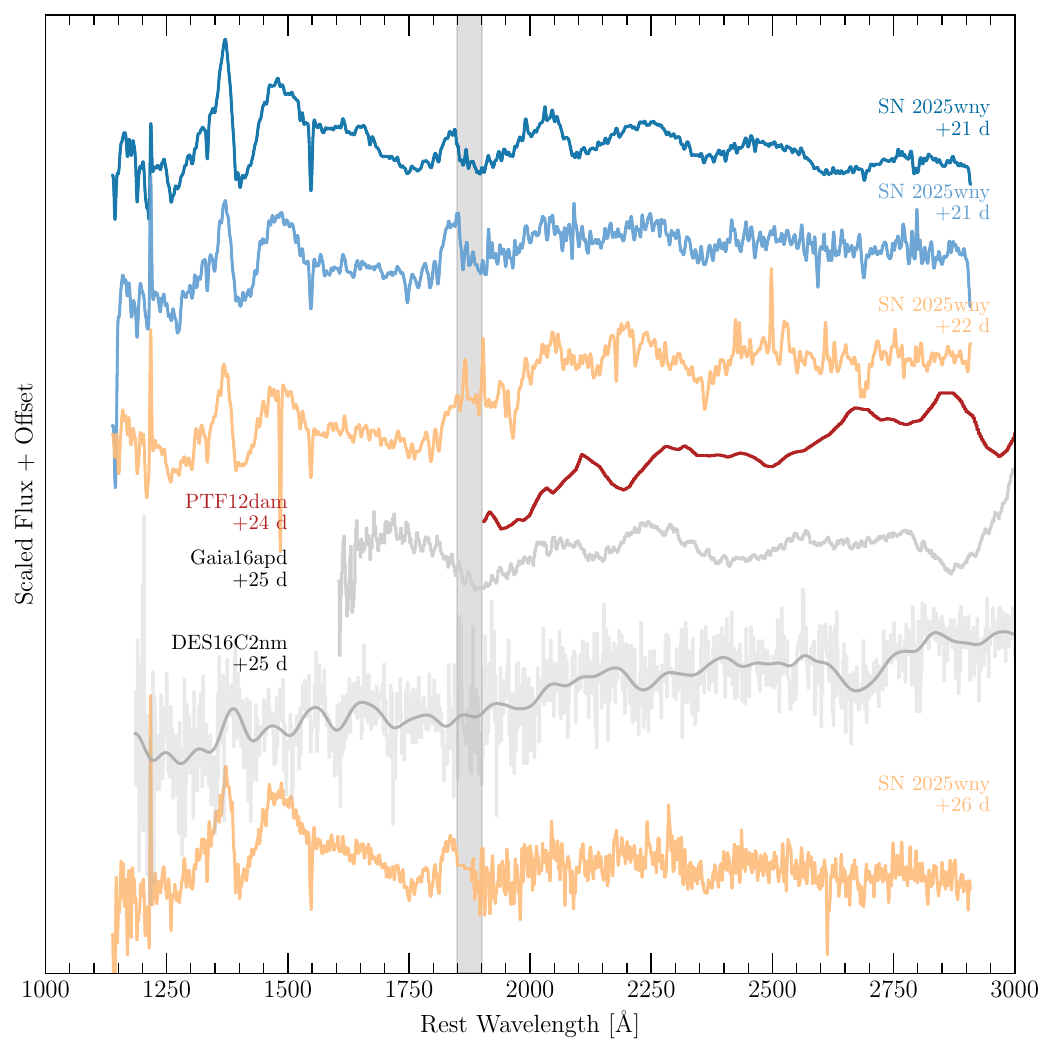}
\caption{Comparison of the rest-frame FUV spectra of H-poor SLSN at $\sim$21--28 days after maximum light. The grey shading denotes the overlap between the blue and red sides of KCWI.}
\label{fig:Spectra_comp_25}
\end{figure*}

Figure~\ref{fig:Spectra_comp_25} presents a comparison between SN~2025wny, DES16C2nm, Gaia16apd, and PTF12dam at $\sim$$+25$~days after maximum light. Only DES16C2nm has a similar wavelength coverage to the spectra of SN~2025wny, but the S/N, which is impressive for a non-lensed source, is too low for confident comparisons. That said, the general spectral shape across the wavelength range is mostly similar, and similar absorption features at $\sim$1400~$\mathrm{\AA}$ may be present. Likewise, Gaia16apd and PTF12dam also show roughly similar spectral shapes, with SN~2025wny and Gaia16apd showing similar line velocities that are faster than those seen in PTF12dam. Finally, the spectra for Gaia16apd, PTF12dam, and DES16C2nm have rising red slopes, in contrast to the mostly flat slope for SN~2025wny. This further demonstrates that SN~2025wny has less line blanketing from iron-group elements than the other three SLSNe-I. Given the similarity with Gaia16apd, the spectra of SN~2025wny likely fall in the lower-metallicity range of typical SLSNe-I spectra.

\subsection{Rest-Frame Optical Spectroscopy}\label{sec:spec_NIR}

\begin{figure*}
\includegraphics[width=\textwidth]{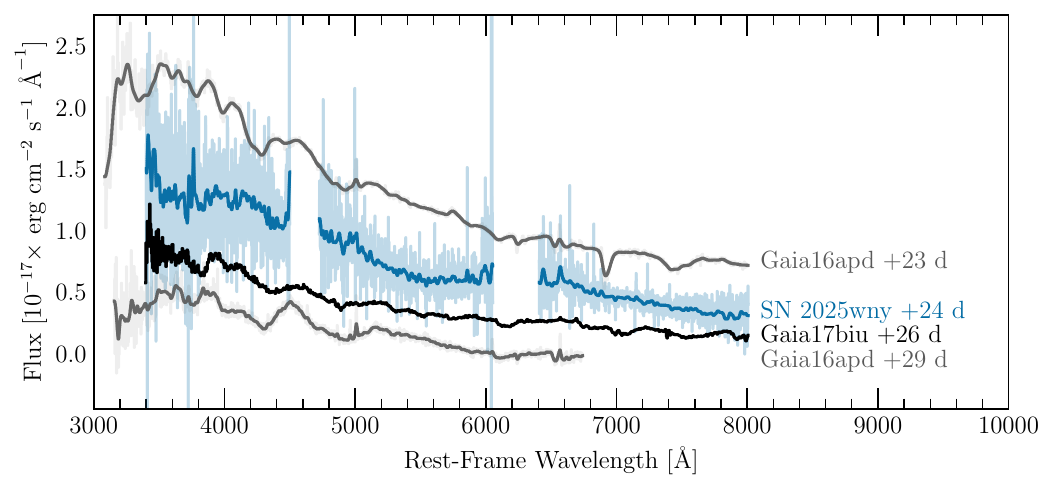}
\caption{Rest-frame optical spectrum taken with NIRES. SN~2025wny is in blue and is compared against Gaia17biu (black; \citealp{Bose18}) and Gaia16apd (grey; \citealp{Kangas17_Gaia16apd, Nicholl17_Gaia16apd}).}
\label{fig:NIR_Spectra}
\end{figure*}

Figure~\ref{fig:NIR_Spectra} shows the rest-frame optical spectrum of SN~2025wny from NIRES at $+27.3$~days after maximum light for image A. \citet{Li26_2025wny} report the absence of \ion{O}{2} in their rest-frame optical spectra from various sources, including \emph{JWST}. Our spectra have a lower S/N, but nonetheless we also do not see the \ion{O}{2} features typically associated with SLSNe-I. 

Figure~\ref{fig:NIR_Spectra} also compares SN~2025wny to a similar-phase optical spectrum of Gaia17biu, a SLSN-I at $z=0.030721$ published in \citet{Bose18}, and Gaia16apd from \citet{Kangas17_Gaia16apd}.\footnote{We note that there appears to be a same-phase optical spectrum reported by \citet{yan2017}, but their Table 3 does not list the spectrum, and it does not appear on WISeREP. The spectrum from \citet{Kangas17_Gaia16apd} is sufficient for our purposes, so we opt for it instead.} The spectra are generally consistent in the optical, with SN~2025wny showing a similarly blue spectral slope. In contrast to our rest-frame UV spectra, our rest-frame optical spectra have lower S/N than those of our comparison objects, preventing a detailed comparison between SN~2025wny and other objects in this wavelength regime. 

\section{Host-Galaxy Spectrum, Outflows, and Intervening ISM}\label{sec:spec_analysis_host}

SNe of all types input kinetic energy into galaxies \citep[e.g.,][]{Simpson2015, Kim15_MomentumInjection, Gutcke21, Keller22, Flury25} and may clear out channels for UV photons (e.g., Ly$\alpha$ and Lyman continuum) to escape \citep[e.g.,][]{Heckman01, heckman2011, Amorin2012, Trebitsch2017, Nelson19, Xu22_CLASSYIII, Carr25, Kong26}.  $Ly\alpha$ escape-fraction measurements in SN hosts, in particular, can help constrain timescales within which, e.g., photoionization heating, stellar winds, and mechanical energy injection from SNe create channels that allow ionizing radiation to escape \citep[e.g.,][]{Kong26, Saldana-Lopez26}.

\begin{figure}
\includegraphics[width=\linewidth]{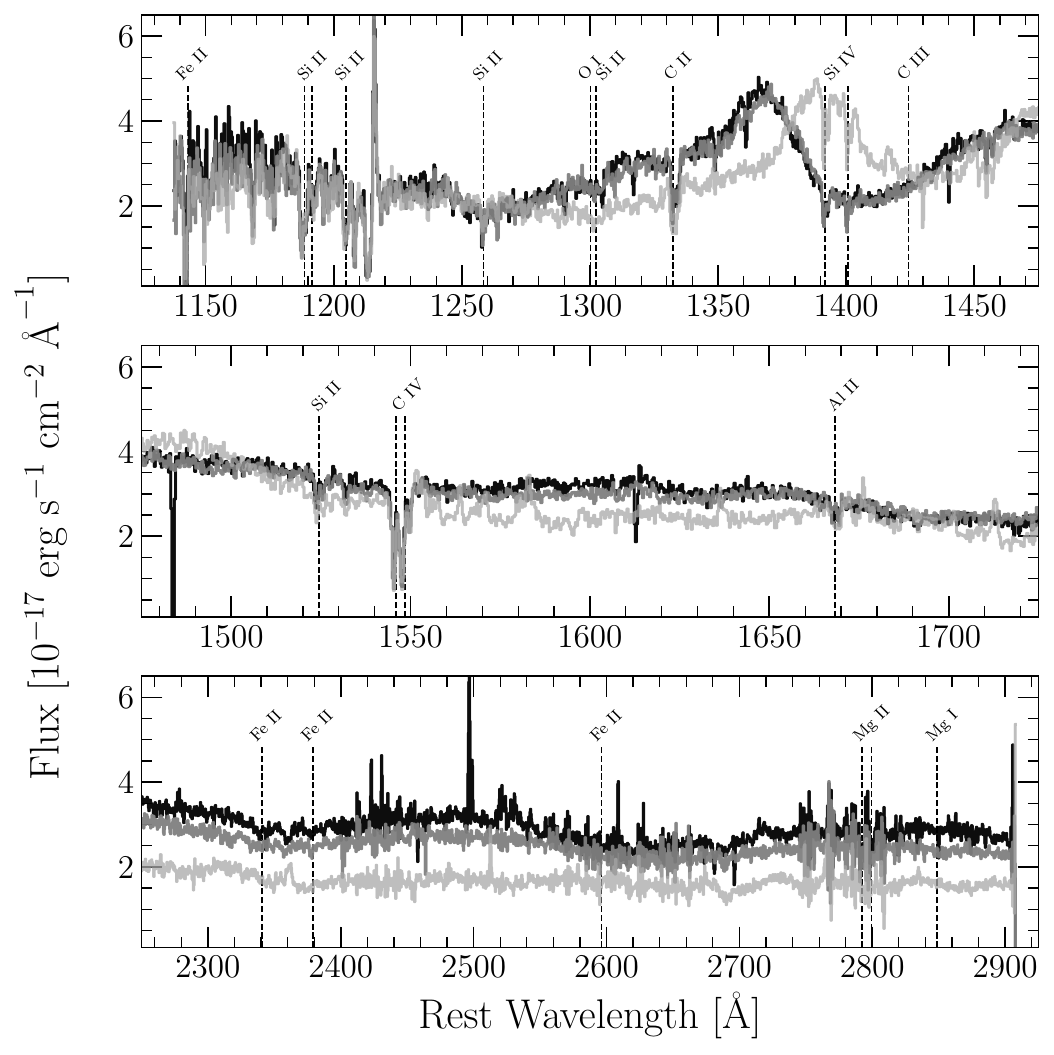}
\caption{Image A spectra of SN~2025wny covering regions with narrow host-galaxy absorption features.  Spectra are at phases of $+18.3$~days, $+21.6$, and $39.4$~days, respectively, from darkest to lightest shading.}
\label{fig:Spectra_Host}
\end{figure}

Narrow interstellar medium (ISM) lines trace this SN processing and feedback. For example, narrow \ion{Si}{2} and \ion{C}{2} absorption trace cold, neutral ISM \citep{Mendez-Hernandez22, Xu23_CLASSY_VI}, whereas \ion{C}{4} and \ion{Si}{4} trace hot gas, winds, and outflows from feedback kinematics \citep{Du2016, Casavecchia24}. The absorption strength of these lines provides column density and neutral-gas covering-fraction measurements \citep{Henry15, Rivera-Thorsen15, Chisholm18_escapefraction}.  Here, we use these features (shown in Figure~\ref{fig:Spectra_Host}) to probe the ISM conditions of SN~2025wny's host galaxy.

In addition to absorption lines, there is also a strong Ly$\alpha$ emission and absorption complex in the host galaxy of SN~2025wny, indicative of star formation \citep{Partridge67}. The origins and profiles of Ly$\alpha$ emission are complex \citep[e.g.,][]{Hansen06}; but, in general, the strength and profile shape of Ly$\alpha$ trace the \ion{H}{1} covering fraction, column density, kinematics, and dust content of the ISM \citep[e.g.,][]{Dijkstra14}.

Extreme Ly$\alpha$ escape can also act as a statistical indicator of Lyman continuum escape. During the epoch of reionization ($5.5<z<12$), these Lyman-leaking galaxies are one of the main sources that contributed to the reionization of surrounding IGM \citep[e.g.,][]{Ouchi09, Robertson13, Robertson15, Finkelstein19, Mason19, Naidu20, Matthee21, Naidu22, Robertson22, Chisholm22}, with SNe potentially playing a role in clearing channels and facilitating this Lyman photon escape \citep[e.g.,][]{Flury25, Kong26}.

To compute an approximate Ly$\alpha$ escape fraction, we integrate the Ly$\alpha$ and H$\alpha$ line fluxes, assuming Case B recombination \citep{Osterbrock89}. This yields, following the method in \citet{Atek08, Atek09}, a Ly$\alpha$ escape fraction of 
\begin{equation}
    f_{esc}^{Ly\alpha}\equiv\frac{F(\mathrm{Ly}\alpha)}{8.7\times F(\mathrm{H}\alpha)}=0.23\pm0.03,
\end{equation}
significantly higher than the average Ly$\alpha$ escape fraction of $\sim$5\% at $z=2.2$ \citep{Hayes2010, Hayes2011, Trainor2015, Naidu22}.

\begin{figure}
\includegraphics[width=\linewidth]{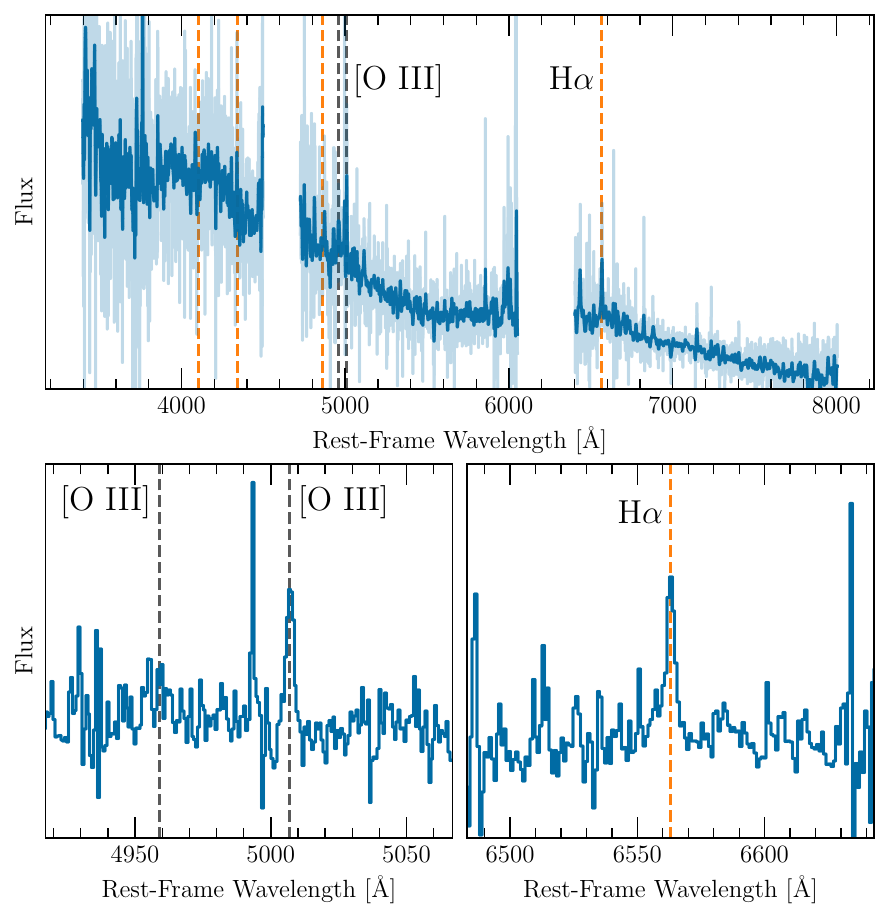}
\caption{\emph{Top:} NIR spectrum of SN~2025wny with Balmer (dashed orange lines) and [\ion{O}{3}] (dashed grey lines) wavelengths  overlaid. We detect no other narrow host-galaxy emission lines in the rest-frame optical. \emph{Bottom Left:} A close-up view of the [\ion{O}{3}]~$\lambda5007$ emission feature. \emph{Bottom Right:} A close up of the H$\alpha$ emission.}
\label{fig:NIR_Galaxy_Lines}
\end{figure}

Lastly, the NIR spectrum shows narrow H$\alpha$ and [\ion{O}{3}]~$\lambda5007$ emission. The previously reported redshift estimates for SN~2025wny, $z=2.008$ and $z=2.010$, do not align well with these narrow lines. We find that $z=2.0155$ explains the rest wavelengths of these two features, as shown in Figure~\ref{fig:NIR_Galaxy_Lines}. This estimate also matches the revision proposed by \citet{Li26_2025wny} to the initial \citet{Taubenberger25} and \citet{Johansson25} estimates. 

\subsection{Narrow ISM Lines}
The spectra of SN~2025wny contain numerous narrow lines from its host galaxy, shown in Figure~\ref{fig:Spectra_Host} with common absorption features labeled. These lines persist throughout the spectral time series and match common galaxy features. These narrow lines constrain the properties of the region where the progenitor of SN~2025wny was located (e.g., its metallicity). Given that the massive-star progenitors of SLSNe live short lives on roughly Myr timescales \citep{Woosley02}, the elemental abundance of material around SN~2025wny reflects the progenitor star's original metallicity pattern. 

\begin{figure*}
    \includegraphics[width=\textwidth]{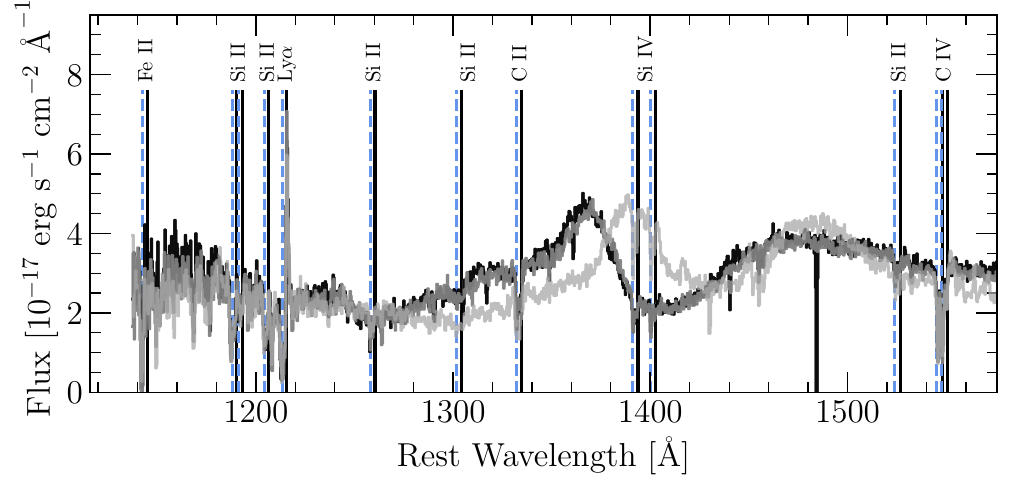}
    \caption{Narrow ISM lines at the rest-frame (solid black) and shifted by $575~\mathrm{km}~\mathrm{s}^{-1}$ (blue dashed).  The rest frame is determined from the narrow emission lines shown in Figure \ref{fig:NIR_Galaxy_Lines}.}
    \label{fig:Spectra_ISM_outflows}
\end{figure*}

We identified several low- and high-ionization features, including \ion{Si}{2}, \ion{C}{2}, \ion{Fe}{2}, \ion{Al}{2}, \ion{C}{4}, \ion{Si}{4}, and \ion{N}{5}. Figure~\ref{fig:Spectra_ISM_outflows} shows some of these lines. Given our redshift determination (which is consistent with recent analysis by \citealt{Li26_2025wny}), we find that the narrow absorption lines require a velocity offset of $\sim$$575~\mathrm{km}~\mathrm{s}^{-1}$. This behavior is rare, with only $\sim$3\% of galaxies analyzed in \citet{Cooper25_outflows} showing such high-velocity outflows, but the velocity is consistent other galaxies with high-velocity outflows \citep[][]{Cooper25_outflows}, suggesting that the host galaxy of SN~2025wny has strong outflows.  Given the presence of one recent massive-star supernova (i.e., SN~2025wny), it is plausible that supernova feedback in the past $\sim$10s of Myr may have contributed to these outflows. Future post-SN~2025wny studies of this galaxy may shed more light on the nature of these outflows. 

\begin{figure*}
    \includegraphics[width=0.24\textwidth]{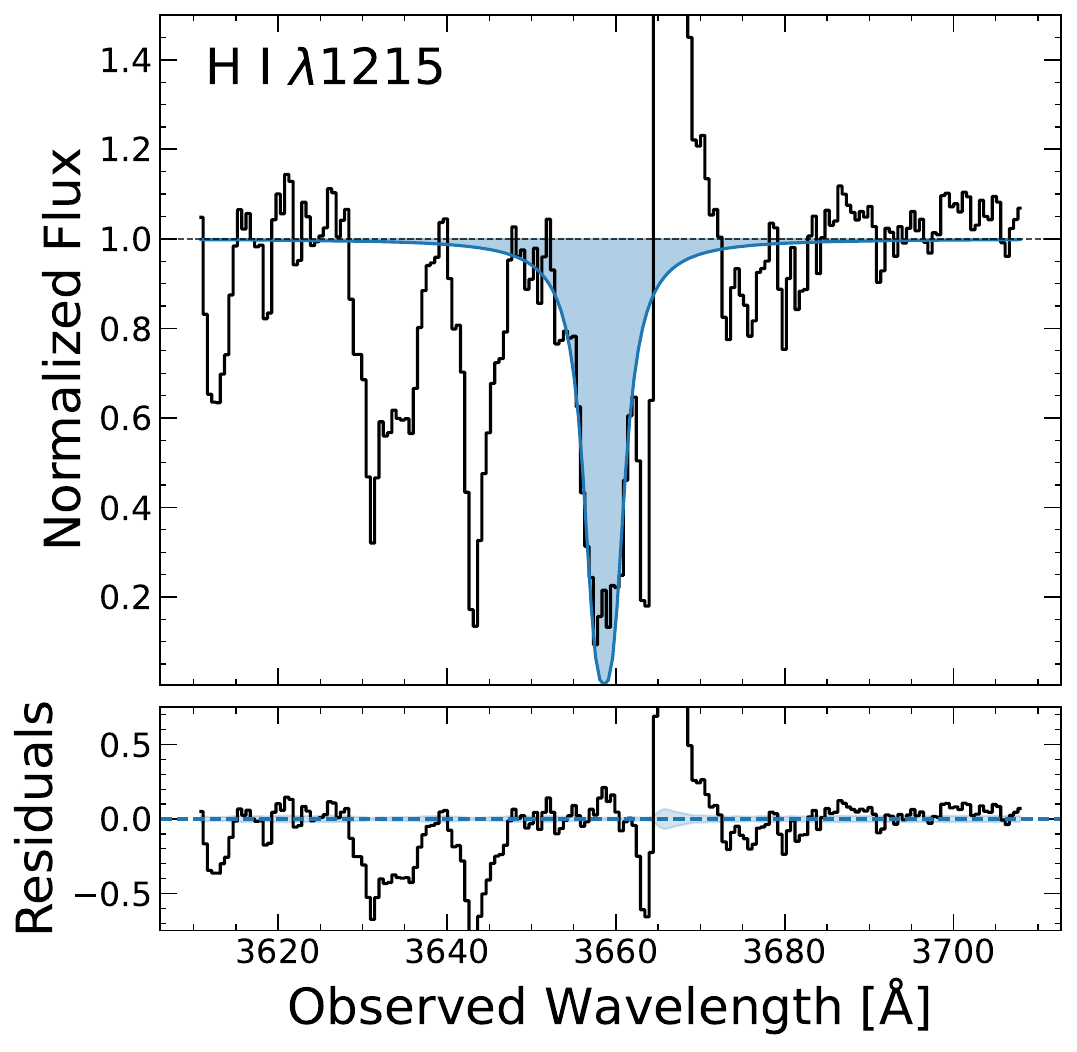} \hfill
    \includegraphics[width=0.24\textwidth]{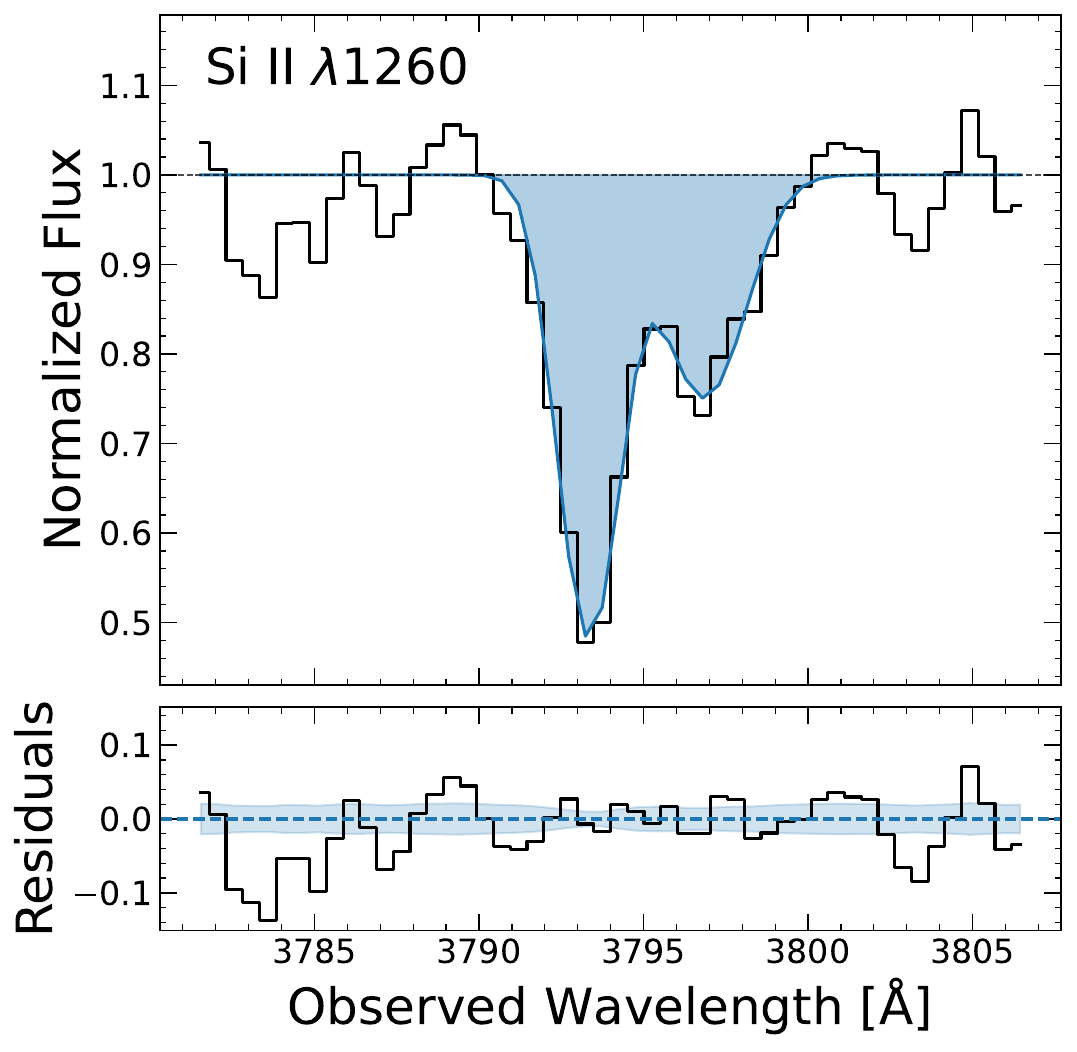} \hfill
    \includegraphics[width=0.24\textwidth]{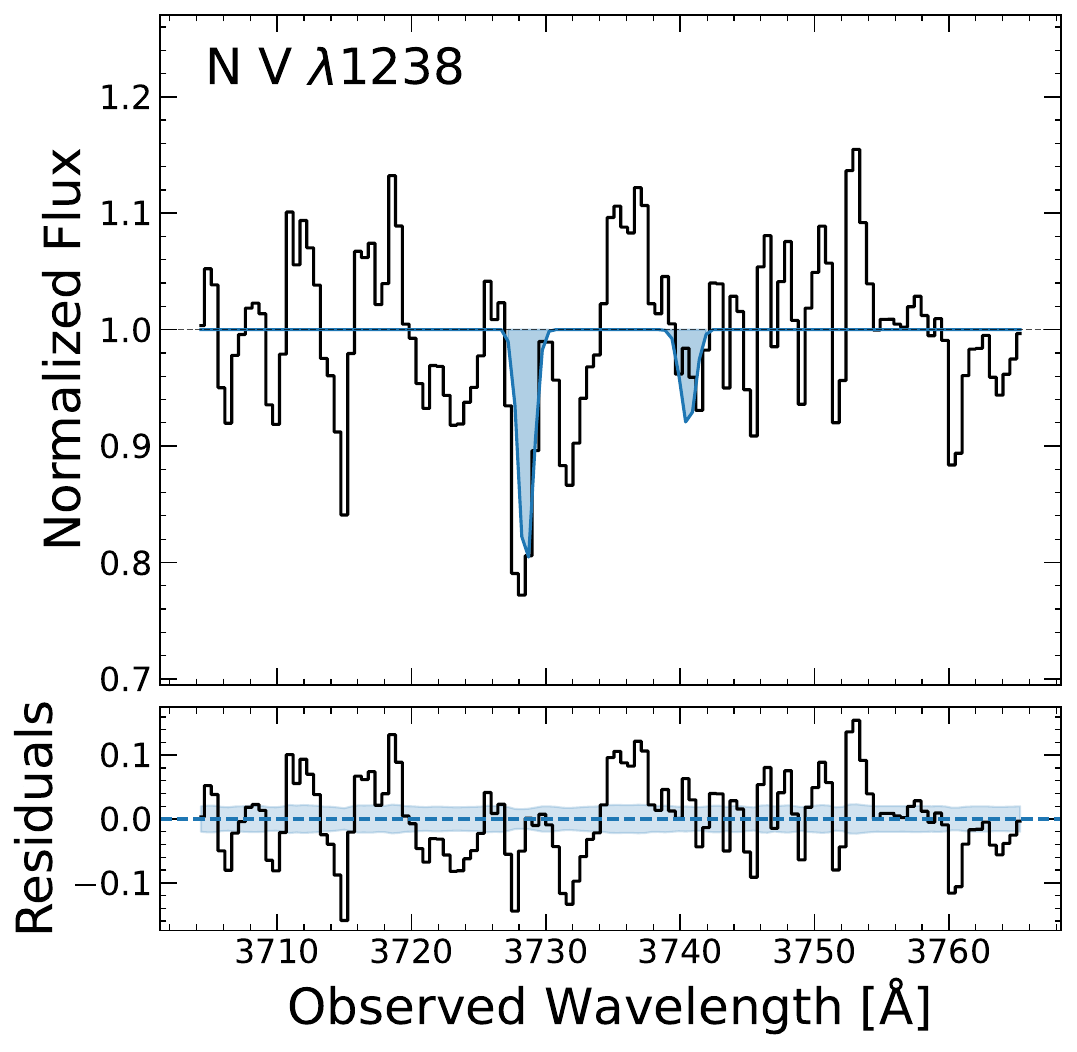} \hfill
    \includegraphics[width=0.24\textwidth]{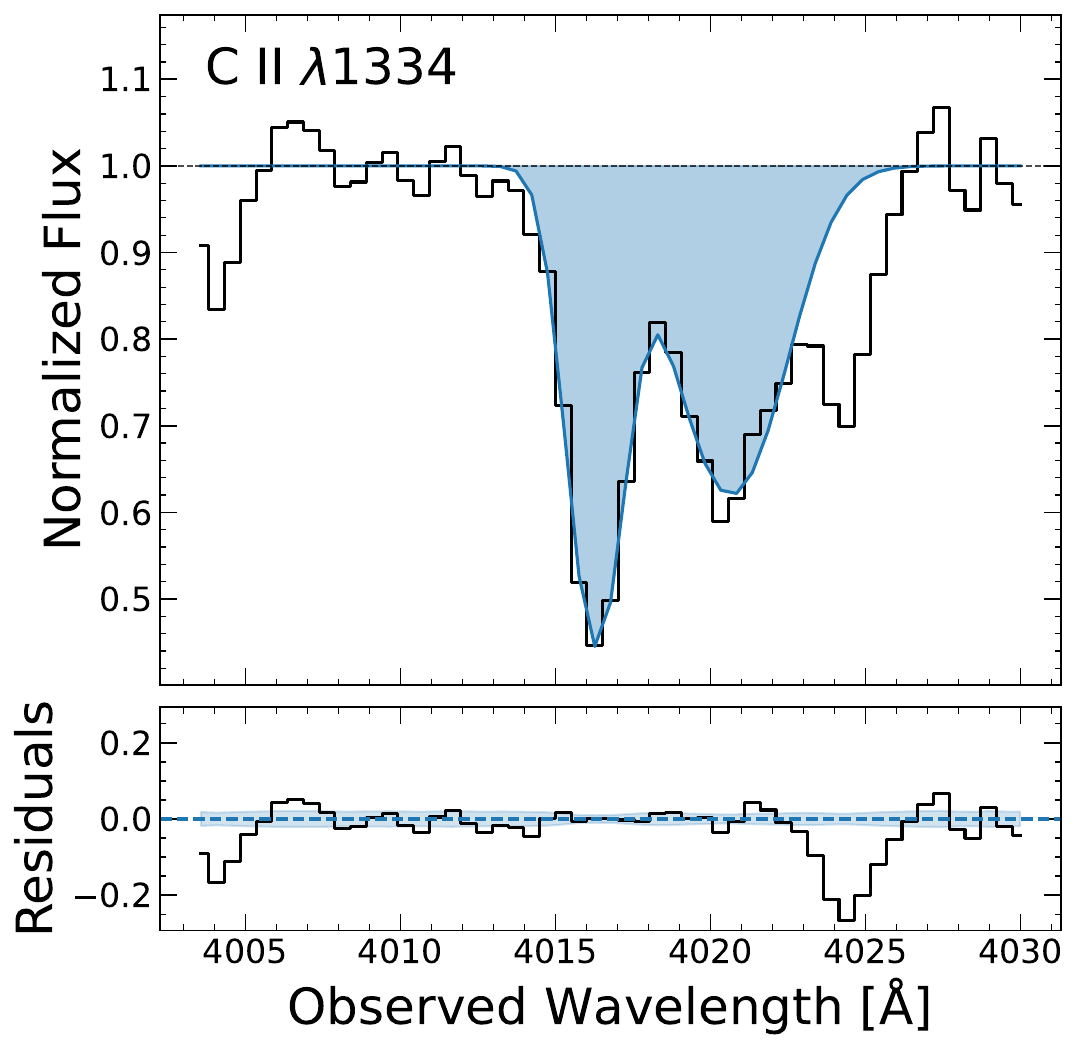} \\
    \includegraphics[width=0.24\textwidth]{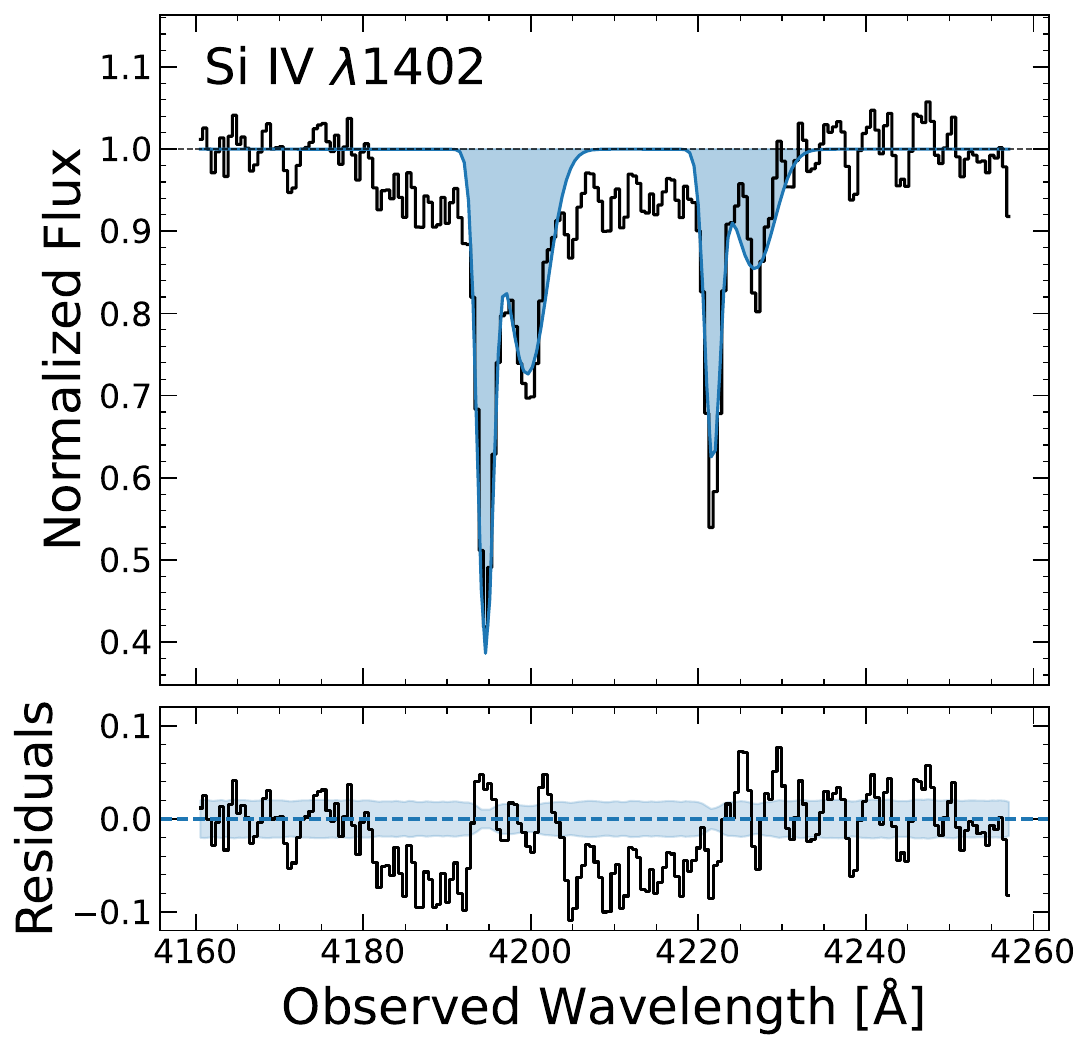} \hfill
    \includegraphics[width=0.24\textwidth]{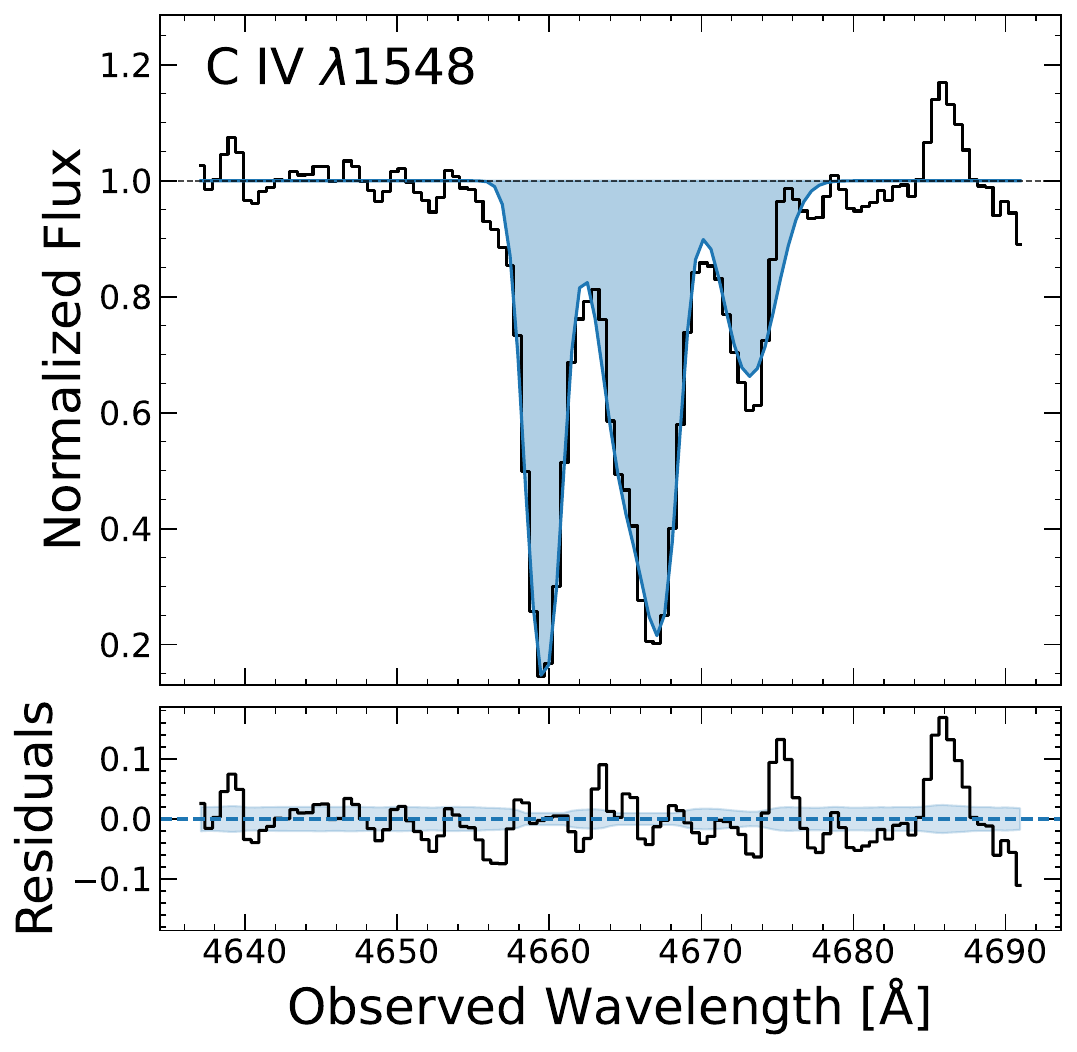} \hfill
    \includegraphics[width=0.24\textwidth]{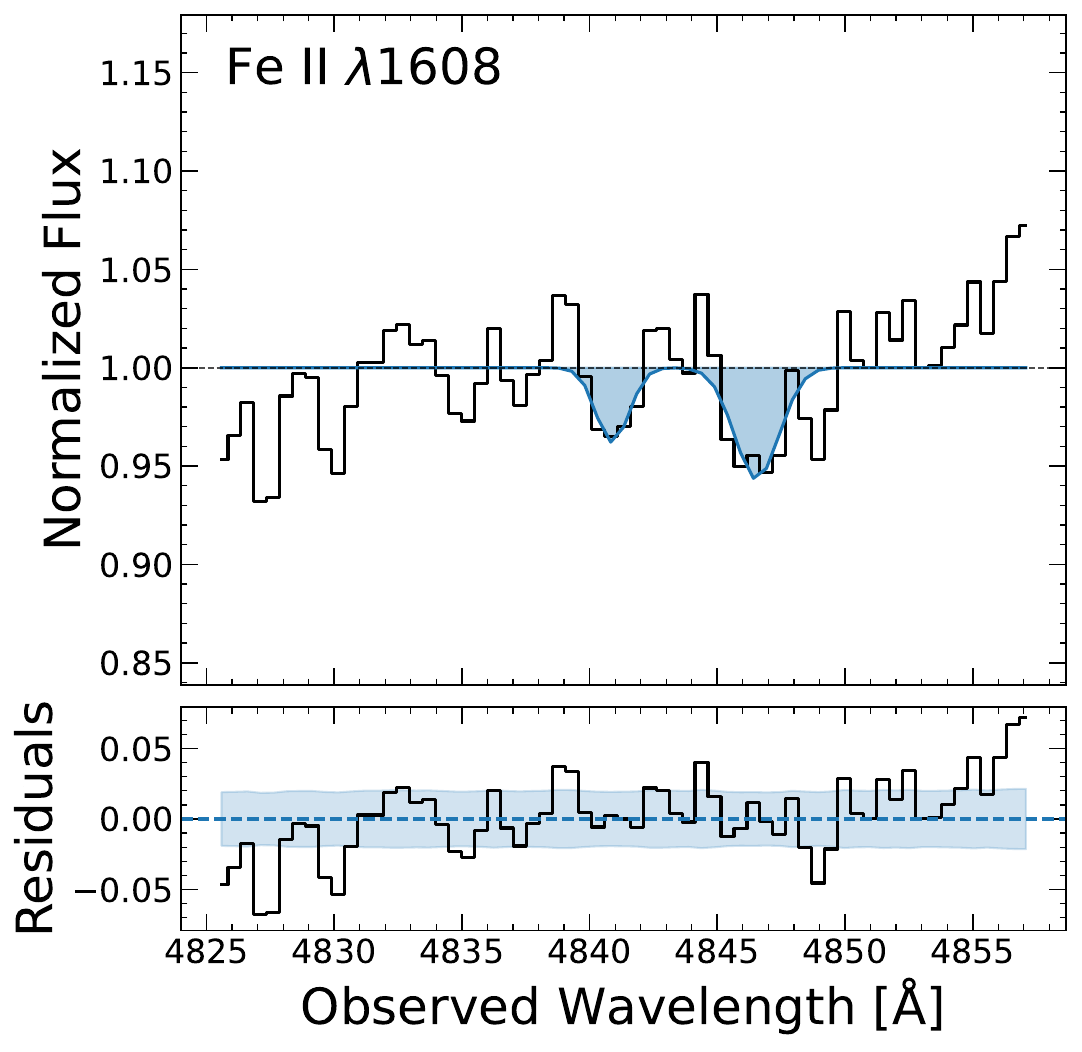} \hfill
    \includegraphics[width=0.24\textwidth]{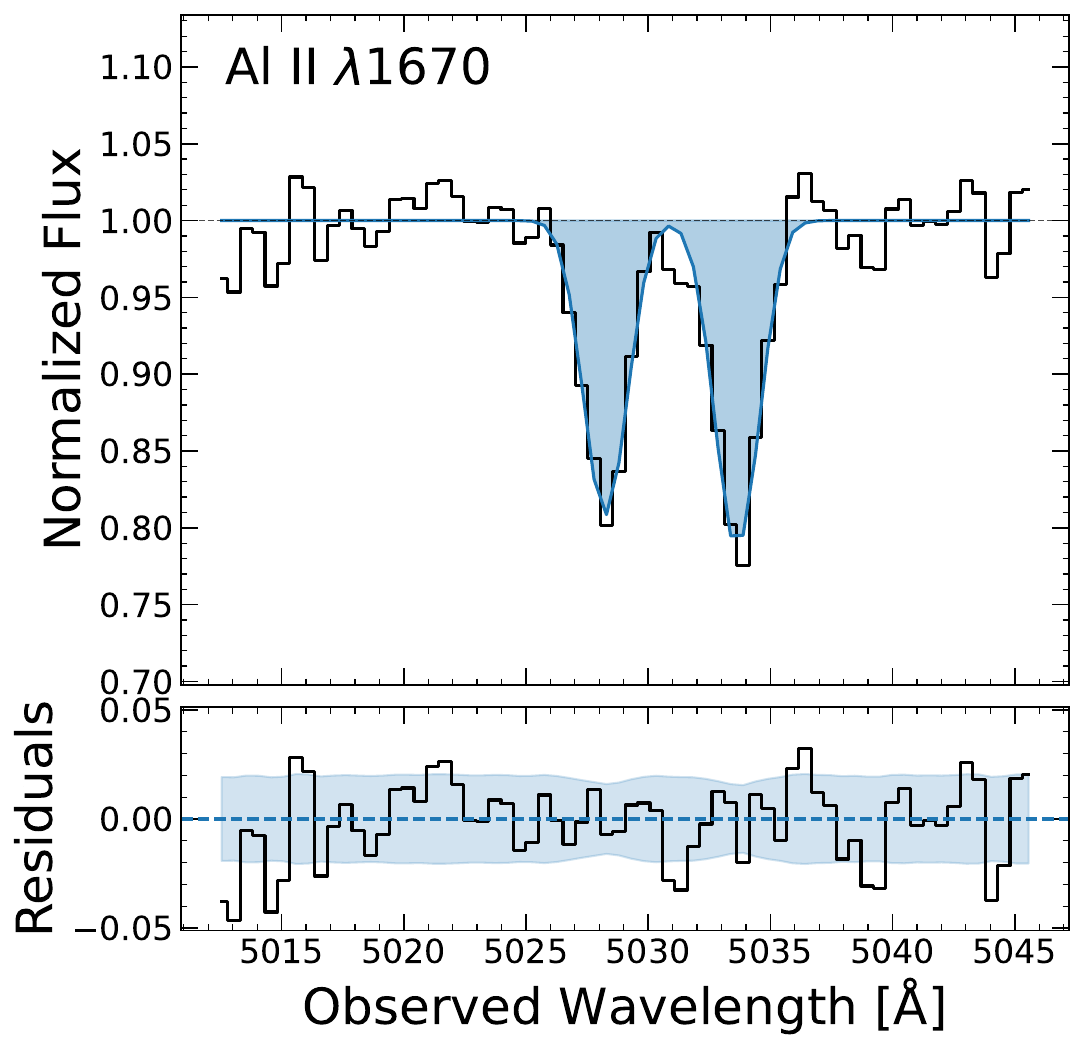} \\
    \caption{Narrow-line fits to the normalized spectrum of SN~2025wny. \ion{C}{4} $\lambda1548$ is a blended doublet.}
    \label{fig:Spectra_ISM_Fits}
\end{figure*}

We measure the column densities of these metal elements using the \texttt{VoigtFit} package \citep{Krogager18}, which simultaneously fits the intrinsic line profiles convolved with the instrumental resolution, thereby determining the velocity ($v$), broadening parameter ($b$), and column densities ($N$) of the considered absorption lines. Figure~\ref{fig:Spectra_ISM_Fits} shows the profile fits to the lines considered in this analysis. Table~\ref{tab:voigtfit} logs the Voigt fit parameters, and Table~\ref{tab:metallicities} logs the final metallicities.

\begin{deluxetable*}{lccc}
\tablecaption{VoigtFit Results for Metal Absorption Lines. \label{tab:voigtfit}}
\tablewidth{0pt}
\tablehead{ \colhead{Ion (Transition)} & \colhead{$v$} & \colhead{$b$} & \colhead{$\log N$} \\
            \colhead{} & \colhead{(km s$^{-1}$)} & \colhead{(km s$^{-1}$)} & \colhead{(cm$^{-2}$)} }
    \startdata
        \ion{H}{1} $\lambda1215$ & $-592.7 \pm 9.0$ & $39.45 \pm 6.78$ & $19.111 \pm 0.094$ \\
        \hline
        \ion{Al}{2} $\lambda1670$ & $-274.3 \pm 6.5$ & $70.60 \pm 10.44$ & - \\
         & $-595.4 \pm 7.2$ & $71.54 \pm 9.27$ & $12.882 \pm 0.036$ \\
        \hline
        \ion{C}{2} $\lambda1334$ & $-270.6 \pm 6.3$ & $84.37 \pm 4.01$ & - \\
         & $-594.5 \pm 2.7$ & $171.63 \pm 11.62$ & $14.774 \pm 0.014$ \\
        \hline
        \ion{Fe}{2} $\lambda1608$ & $-228.8 \pm 24.3$ & $54.59 \pm 50.87$ & - \\
         & $-578.2 \pm 34.3$ & $79.40 \pm 37.74$ & $13.748 \pm 0.150$ \\
        \hline
        \ion{Si}{2} $\lambda1260$ & $-313.6 \pm 10.1$ & $88.06 \pm 4.94$ & - \\
         & $-587.4 \pm 3.3$ & $130.05 \pm 15.93$ & $13.574 \pm 0.020$ \\
        \hline
        \ion{C}{4} $\lambda\lambda1548,1550$ & $-207.7 \pm 5.5$ & $64.65 \pm 2.85$ & - \\
         & $-570.9 \pm 1.7$ & $113.01 \pm 7.75$ & $14.876 \pm 0.018$ \\
        \hline
        \ion{N}{5} $\lambda\lambda1238,1242$ & $-592.4 \pm 9.0$ & $76.33 \pm 24.82$ & $13.847 \pm 0.062$ \\
        \hline
        \ion{Si}{4} $\lambda\lambda1393,1402$ & $-236.2 \pm 14.1$ & $63.94 \pm 5.92$ &  \\
         & $-591.7 \pm 3.1$ & $227.02 \pm 22.55$ & $14.165 \pm 0.019$
    \enddata
\tablecomments{Velocities are relative to the systemic redshift $z = 2.0155$. Each row without a
repeated ion label corresponds to an additional velocity component of the same
transition, fit simultaneously with \texttt{VoigtFit} \citep{Krogager18}.}
\end{deluxetable*}

\begin{deluxetable*}{lccc}
\tablecaption{Metallicities and Total Column Densities. \label{tab:metallicities}}
\tablewidth{0pt}
\tablehead{ \colhead{Ion} & \colhead{$\log N_{\rm tot}$} & \colhead{[X/H]$_{\rm app}$} & \colhead{[X/H]$_{\rm corr}$} \\
            \colhead{} & \colhead{(cm$^{-2}$)} & \colhead{} & \colhead{} }
    \startdata
        \ion{H}{1}  & $19.112 \pm 0.093$ & \nodata            & \nodata \\
        \ion{Al}{2} & $12.884 \pm 0.036$ & $-0.659 \pm 0.102$ & $-1.962 \pm 0.102$ \\
        \ion{C}{2}  & $14.774 \pm 0.015$ & $-0.798 \pm 0.102$ & $-1.481 \pm 0.102$ \\
        \ion{Fe}{2} & $13.779 \pm 0.157$ & $-0.795 \pm 0.183$ & $-1.080 \pm 0.183$ \\
        \ion{Si}{2} & $13.574 \pm 0.020$ & $-1.049 \pm 0.096$ & $-2.048 \pm 0.096$ \\
        \ion{C}{4}  & $14.876 \pm 0.018$ & $-0.696 \pm 0.101$ & $-2.275 \pm 0.101$ \\
        \ion{N}{5}  & $13.847 \pm 0.061$ & $-1.097 \pm 0.131$ & $-2.061 \pm 0.131$ \\
        \ion{Si}{4} & $14.165 \pm 0.019$ & $-0.457 \pm 0.093$ & $-2.070 \pm 0.093$
    \enddata
\tablecomments{Total column densities ($\log N_{\rm tot}$) are summed over all velocity components before
dust depletion correction. Apparent metallicities ([X/H]$_{\rm app}$) are computed relative to solar
abundances using the \ion{H}{1} column density. Corrected metallicities ([X/H]$_{\rm corr}$) apply
ionization correction factors (ICFs) derived from our \texttt{CLOUDY} model to account for the hidden ionization states of both the metals and hydrogen.}
\end{deluxetable*}

For Ly$\alpha$, we fit the absorption component with \texttt{VoigtFit}, with a single absorption component ($v_I = -592.7 \pm 9.0$~km s$^{-1}$), despite the spectrum showing a narrow absorption at red wavelengths, which is consistent in velocity with the faint components identified in metal lines (see below). We obtain a total (both components) hydrogen column density of $\log_{10}\left({N(\mathrm{H})}\right) = 19.11 \pm 0.09$. The column densities measured in the spectrum of SN~2025wny are reported in Table~\ref{tab:voigtfit}. For metal lines we provided an integrated column density obtained from the summed metal column densities of each single absorption.

The simultaneous detection of both low-ionization (\ion{Si}{2}) and high-ionization (\ion{Si}{4}) absorption features along the line of sight implies that the observed gas does not originate from a single, uniform neutral phase. This is also supported by the relatively low neutral hydrogen column density and high escape fraction, implying that the gas is unlikely to be fully neutral \citep[e.g.,][]{Saccardi23}. Instead, it probes a complex, multi-phase interstellar medium (ISM) or a boundary layer subjected to a significant ionizing radiation field \citep[e.g.,][]{Savage96, Howk99}. Under these conditions, the usual approximation that the total gas-phase abundance is dominated by the singly ionized state is physically invalid because a substantial fraction of the element is expected to be ionized into higher states.

To account for ionization fractions and derive an accurate absolute abundance, we modeled the cloud's ionization state using photoionization equilibrium frameworks with grids generated by the \texttt{CLOUDY} software package \citep{2013RMxAA..49..137F, Ferland17_CLOUDY, Cloudy_2025}, version 25.0. Focusing only on elements for which we have a low- and high-ionization measurement (i.e., Si), we find that the total measured column densities along our line of sight are $\log N($\ion{Si}{2}$) = 13.57 \pm 0.02$ and $\log N($\ion{Si}{4}$) = 14.16 \pm 0.02$, yielding an observed high-to-low ion ratio of $\log [N(\text{Si IV}) / N($\ion{Si}{2}$)] = 0.59 \pm 0.03$. Because the ionization potential required to convert \ion{Si}{3} to \ion{Si}{4} is relatively high ($33.49\text{ eV}$), environments capable of generating such an elevated \ion{Si}{4} to \ion{Si}{2} ratio inevitably produce an ionization bottleneck where the intermediate ion, \ion{Si}{3}, becomes the dominant gas-phase reservoir \citep{Prochaska02_II}.

To map our observed constraints onto the underlying physical ion populations robustly and to track non-linear error propagation across logarithmic scales, we used a cubic-spline interpolation bound to a photoionization equilibrium grid and evaluated it over 10,000 Monte Carlo realizations. Our simulations reveal that approximately $60\%$ of the total Si column density is sequestered in the unobserved \ion{Si}{3} state, while \ion{Si}{2} and \ion{Si}{4} constitute roughly $9\%$ and $36\%$ of the total budget, respectively. The resulting derived column densities are: $\log N($\ion{Si}{3}$) = 14.46 \pm 0.02$ and $\log N(\mathrm{Si}_{\mathrm{total}}) = 14.68 \pm 0.02$. Crucially, if the ionization correction factor (ICF) were omitted and the abundance derived solely from $N($\ion{Si}{2}$)$, the absolute silicon column density would be systematically underestimated by $1.04$~dex (or a factor of $2.21$ if combining both observed \ion{Si}{2} and \ion{Si}{4}). Furthermore, because hydrogen is highly ionized in this environment ($\Delta \log N(\mathrm{H}) \approx 2.11$~dex), applying the full multi-phase ionization corrections systematically alters the derived absolute gas-phase metallicities. Specifically, scaling the corrected total silicon against only the neutral \ion{H}{1} column yields an artificially inflated, near-solar metallicity of $[\mathrm{Si/H}] \approx 0.06$. Properly accounting for the dominant, unseen reservoir of ionized hydrogen (\ion{H}{2}) resolves this observational bias, revealing a heavily sub-solar intrinsic metallicity of $[\mathrm{Si/H}] = -2.05 \pm 0.10$. Table~\ref{tab:metallicities} compiles the final metallicities and column densities measured by the narrow absorption features.

\begin{figure}
\includegraphics[width=\linewidth]{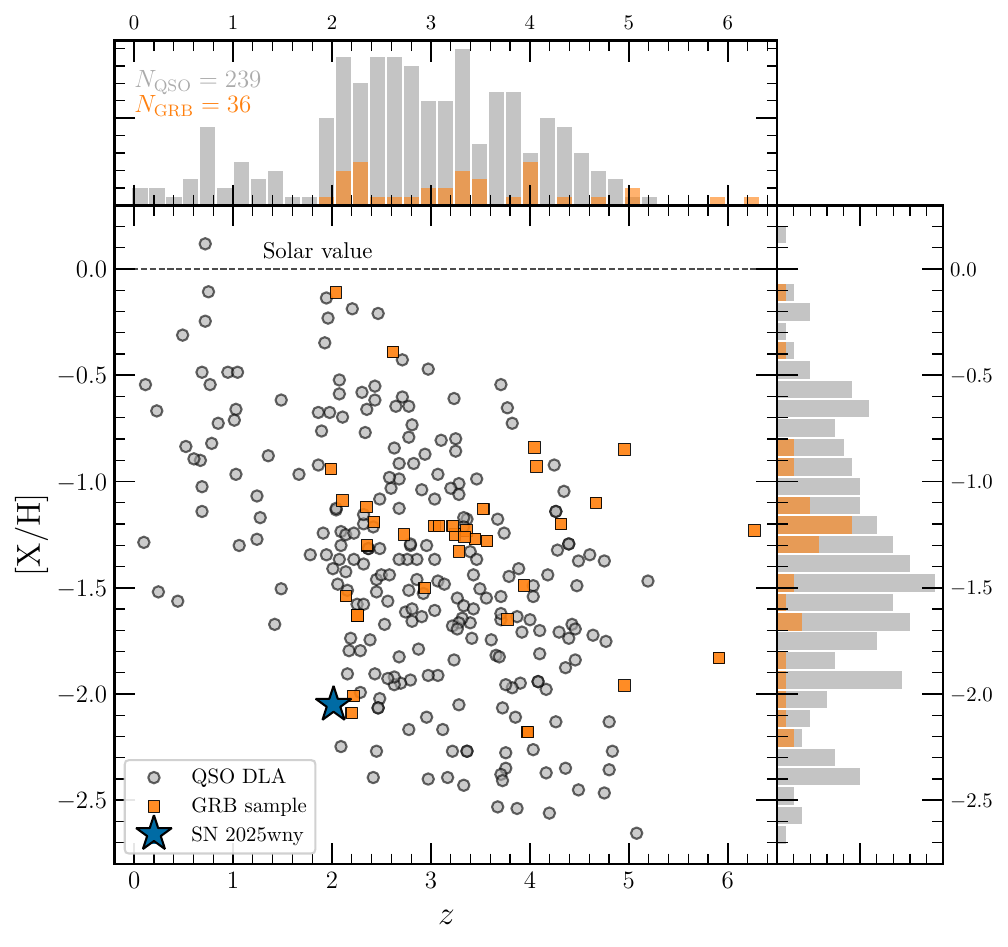}
\caption{Host-galaxy metallicity inferred from narrow-line absorption features versus redshift for SN~2025wny (from \ion{Si}{2} measurements; blue star), quasars (grey circles) and gamma-ray bursts (orange squares). The metallicity is inferred mostly from \ion{Si}{2} lines, however, for some sources \ion{Fe}{2} is used instead.}
\label{fig:metallicities}
\end{figure}

Finally, the metal column densities obtained above do not represent the intrinsic column densities of the local ISM, as they are affected by dust depletion \citep[where not all the metals are in the gas phase and some are part of dust grains, see, e.g.,][]{Konstantopoulou22}. We therefore apply dust-depletion corrections to derive intrinsic abundances, using the prescriptions in \citet{DeCia16} and \citet{DeCia18}. In these works, dust depletion corrections depend on the [Zn/Fe] values measured in spectra. Because the SN~2025wny spectra do not directly measure [Zn/Fe], we rely on the relative abundance [Si/Fe]. Integrating the singly-ionized column densities across all velocity components yields a total apparent relative abundance of $\mathrm{[Si/Fe]}_{\mathrm{obs}} = -0.25 \pm 0.18$. This value corresponds to an equivalent expected [Zn/Fe] value using the linear fits from \citet{DeCia16} and \citet{DeCia18}: 

\begin{equation}
    \mathrm{\left[\frac{\mathrm{Zn}}{\mathrm{Fe}} \right]}_{\mathrm{exp}} = \frac{\left[\frac{\mathrm{Si}}{\mathrm{Fe}} \right]_\mathrm{obs} - A1_\mathrm{Si} }{B1_\mathrm{Si} + 1}
\end{equation}

where $A1_\mathrm{Si}$ and $B1_\mathrm{Si}$ are the empirically derived fit coefficients for Si. Given our sub-solar $\mathrm{[Si/Fe]}_{\mathrm{obs}}$ measurement, this relation yields $\mathrm{[Zn/Fe]} \lesssim 0$. Physically, a negative or near-zero [Zn/Fe] indicates that dust depletion along this highly ionized line of sight is negligible, meaning the massive $+1.04$~dex ionization correction derived from our \texttt{CLOUDY} model heavily dominates over any minor dust depletion effects. Consequently, our final intrinsic metallicity remains $\mathrm{[Si/H]}_{\mathrm{corr}} = -2.05 \pm 0.10$. Given the lack of a comparison sample of cold-gas metallicities in high-redshift SLSN environments, we compare our measurement with the metallicity distribution derived from absorption studies of gamma-ray burst (GRB) and QSO-Damped Lyman $\alpha$ (DLA) systems \citep{Rafelski12, Thone13, Bolmer19, Saccardi23} shown in Figure \ref{fig:metallicities}. In this comparison, [X/H] is primarily derived from the ionic species \ion{S}{2}, \ion{Zn}{2}, \ion{Si}{2}.

We find that SN~2025wny exploded in a sub-solar-metallicity environment, broadly consistent with metallicities derived from GRB and quasar narrow lines. The rest-frame optical spectrum lacks sufficient lines to directly measure [O/H]; however, adopting the mean low-ion abundance (Al, C, Fe, and Si) as a proxy for O, we find $12+\log(\mathrm{O/H})=7.9\pm0.2$ on the solar scale of \citet{Asplund21}, uncorrected for dust depletion or ionization. 
% This value is 0.4~dex lower than the Qin et al.\ in preparation value quoted by \citet{Li26_2025wny}. % Not public yet, commenting out until on arxiv. 
Using the NIR spectrum H$\alpha$ line and the relationship in \citet{Kennicutt94}, we find a star formation rate of $\sim$1.4~$\mathrm{M}_\odot$/yr.

\section{Conclusions}\label{sec:conclusions}

We present extensive rest-frame UV observations of the strongly gravitationally lensed SLSN 2025wny, including photometry from YSE and ZTF, and the first FUV spectral time series of an SLSN-I spanning the first two months after maximum light. 

The light curves near maximum light are well fit by the \texttt{slsnni} model in MOSFiT, which models an engine-driven explosion with additional energy input from radioactive $^{56}\mathrm{Ni}$. SN~2025wny has magnetar parameters consistent with local-universe samples of SLSNe \citep{Nicholl17, gomez2024, Hinkle25_SwiftFixII}; however, our model cannot explain the observed plateau $\sim$20~days after maximum light. Similar to \citet{Li26_2025wny}, we posit that this plateau may arise from additional CSM interaction or central-engine feedback from fallback accretion. 

The rest-frame UV spectra of SN~2025wny resemble those of the few lower-redshift SLSNe with rest-frame UV spectroscopy, with similar spectroscopic features. Population comparisons are tentative at best, given the paucity of spectra with the same wavelength coverage as those presented here. The rest-frame optical spectrum taken $\sim$1~month after image A maximum light lacks the normally conspicuous \ion{O}{2} features typically associated with SLSNe-I and resembles the optical spectrum of Gaia17biu, one of the lowest-redshift SLSNe to date that exploded in a metal-rich galaxy \citep{Bose18}.

Narrow ISM lines and Ly$\alpha$ emission superimposed on the SN~2025wny spectra enable analysis of its host galaxy. Using \texttt{CLOUDY} to model the ionization states of the narrow lines, we find that SN~2025wny exploded in a sub-solar metallicity galaxy, and its host-galaxy metallicity is consistent with QSO and GRB host galaxies at similar redshifts. Additionally, we find that the host-galaxy of SN~2025wny has an exceptionally high escape fraction ($f_{esc}^{Ly\alpha}=0.23\pm0.03$) and high-velocity outflows of $\sim$$575~{\rm km~s^{-1}}$.

Such rare, long-term, high-S/N UV data will provide a benchmark for improving empirical SLSNe SED models needed to calculate $K$-corrections for the abundance of high-redshift SLSNe expected from \emph{Roman} and LSST, both of which are on the precipice of transforming time-domain astronomy. Furthermore, as strong lensing searches \citep[e.g.,][]{huang20201, huang20211, storfer2024} increase the number of known lenses, the number of strongly lensed supernovae will increase, creating a new avenue to study statistical rest-frame UV spectroscopic properties of SNe. 

In summary, SN~2025wny is relatively typical in magnetar model explosion parameters and luminosity. The late-time energy injection that forms a light-curve plateau is slightly atypical compared to normal SLSNe-I but attributable to a typical process invoked for SLSNe-I such as CSM interaction or fallback accretion. The host-galaxy metallicity falls within the range seen for other SLSN host galaxies. 

The high-velocity, $\sim$$ $$575~\rm{km~s^{-1}}$ outflows seen in the host galaxy of SN~2025wny are particularly interesting given recent results from \citet{Saldana-Lopez26}, who find that SLSN hosts have maximum outflow velocities of $34$--$104~{\rm km~s^{-1}}$ from a sample of 6 SLSNe with narrow ISM absorption; their finding is consistent with the expectation that outflows are stellar-wind dominated, as SN feedback has not yet had time to influence the ISM in these hosts.  In this context, our results may suggest that some SLSN hosts have higher-than-expected wind-driven feedback, more extended (less bursty) star-formation histories, or shorter timescales for SN feedback to influence their surrounding ISM than anticipated.

Future samples from \emph{Roman} and LSST will drastically increase the sample of SLSNe-I in the early universe. Both at cosmic noon and the epoch of reionization, the progenitor and host-galaxy metallicities of these explosions should be lower than in the local universe, enabling tests of whether post-peak CSM or fallback accretion contributions to the light curve become more prevalent in SLSNe-I. Such objects will also test whether their host galaxies show unique feedback properties (i.e., peculiar outflow velocities or narrow-line ISM properties) tied to SLSN-I explosions, possibly offering new insights into the role of mechanical feedback from metal-poor massive stars. 

\begin{acknowledgments}
We thank Jennifer van Saders and Daichi Hiramatsu for helpful discussions. 
W.B.H. acknowledges support from the National Science Foundation Graduate Research Fellowship Program under Grant No. 2236415. 
D.O.J. acknowledges support from NSF grants AST-2407632, AST-2429450, and AST-2510993, NASA grant 80NSSC24M0023, and HST/JWST grants HST-GO-17128.028 and JWST-GO-05324.031, awarded by the Space Telescope Science Institute (STScI), which is operated by the Association of Universities for Research in Astronomy, Inc., for NASA, under contract NAS5-26555.
C.J.S. and E.A.M. are supported by NASA Euclid General Investigator Program Grant No. 80NSSC25K7735 and NASA Roman ROSES Grant No. 80NSSC24K0353.
L.I. acknowledges financial support from the INAF Data Grant Program {\it Multi-wavelength and multi-messenger analysis of relativistic supernovae}.
D.R.A.D. acknowledges support from the Lamat Institute \citep{2025NatAs...9.1770Q} and UC Santa Cruz, supported by the Heising-Simons Foundation. 
K.S.M. and M.G. are supported by the European Union’s Horizon 2020 research and innovation programme under European Research Council Grant Agreement No. 101002652 (BayeSN; PI K. Mandel).
J.H. was supported by a research grant (VIL54489) from VILLUM FONDEN. 
C.T.M. thanks the LSST-DA Data Science Fellowship Program, which is funded by LSST-DA, the Brinson Foundation, the WoodNext Foundation, the Research Corporation for Science Advancement Foundation, and Kevin Wells; his participation in the program has benefited this work. 
G.N. is funded by NSF CAREER grant AST-2239364, supported in-part by funding from Charles Simonyi. G.N. also gratefully acknowledges NSF support from OAC-2311355, AST-2432428, as well as AST-2421845 and funding from the Simons Foundation for the NSF-Simons SkAI Institute. 
G.N. is also supported by the DOE through the Department of Physics at the University of Illinois, Urbana-Champaign.
Pan-STARRS is a project of the Institute for Astronomy of the University of Hawai\okina i, and is supported by the NASA SSO Near Earth Observation Program under grants 80NSSC18K0971, NNX14AM74G, NNX12AR65G, NNX13AQ47G, NNX08AR22G, 80NSSC21K1572 and by the State of Hawai\okina i.

Some of the data presented herein were obtained at Keck Observatory, which is a private 501(c)3 non-profit organization operated as a scientific partnership among the California Institute of Technology, the University of California, and the National Aeronautics and Space Administration. The Observatory was made possible by the generous financial support of the W. M. Keck Foundation. 

The authors wish to recognize and acknowledge the very significant cultural role and reverence that the summit of Maunakea has always had within the Native Hawaiian community. We are most fortunate to have the opportunity to conduct observations from this mountain.

This research has made use of the Keck Observatory Archive (KOA), which is operated by the W. M. Keck Observatory and the NASA Exoplanet Science Institute (NExScI), under contract with the National Aeronautics and Space Administration.
\end{acknowledgments}

\facilities{Keck-II(KCWI, NIRES), PS1}

\software{NumPy \citep{numpy}, 
          SciPy \citep{scipy}, 
          AstroPy \citep{2013A&A...558A..33A, 2018AJ....156..123A, 2022ApJ...935..167A}, 
          Matplotlib \citep{matplotlib},
          Pandas \citep{pandas},
          MOSFiT \citep{Guillochon18_MOSFiT},
          VoigtFit \citep{Krogager18},
          CLOUDY \citep{2013RMxAA..49..137F, Ferland17_CLOUDY, Cloudy_2025},
          lenstronomy \citep{birrer2018},
          GLEE \citep{suyu20101, suyu20121},
          emcee \citep{Foreman-Mackey13}
          }

\bibliography{sample701}{}
\bibliographystyle{aasjournalv7}

\begin{figure*}
        \centering
        \includegraphics[width=\textwidth]{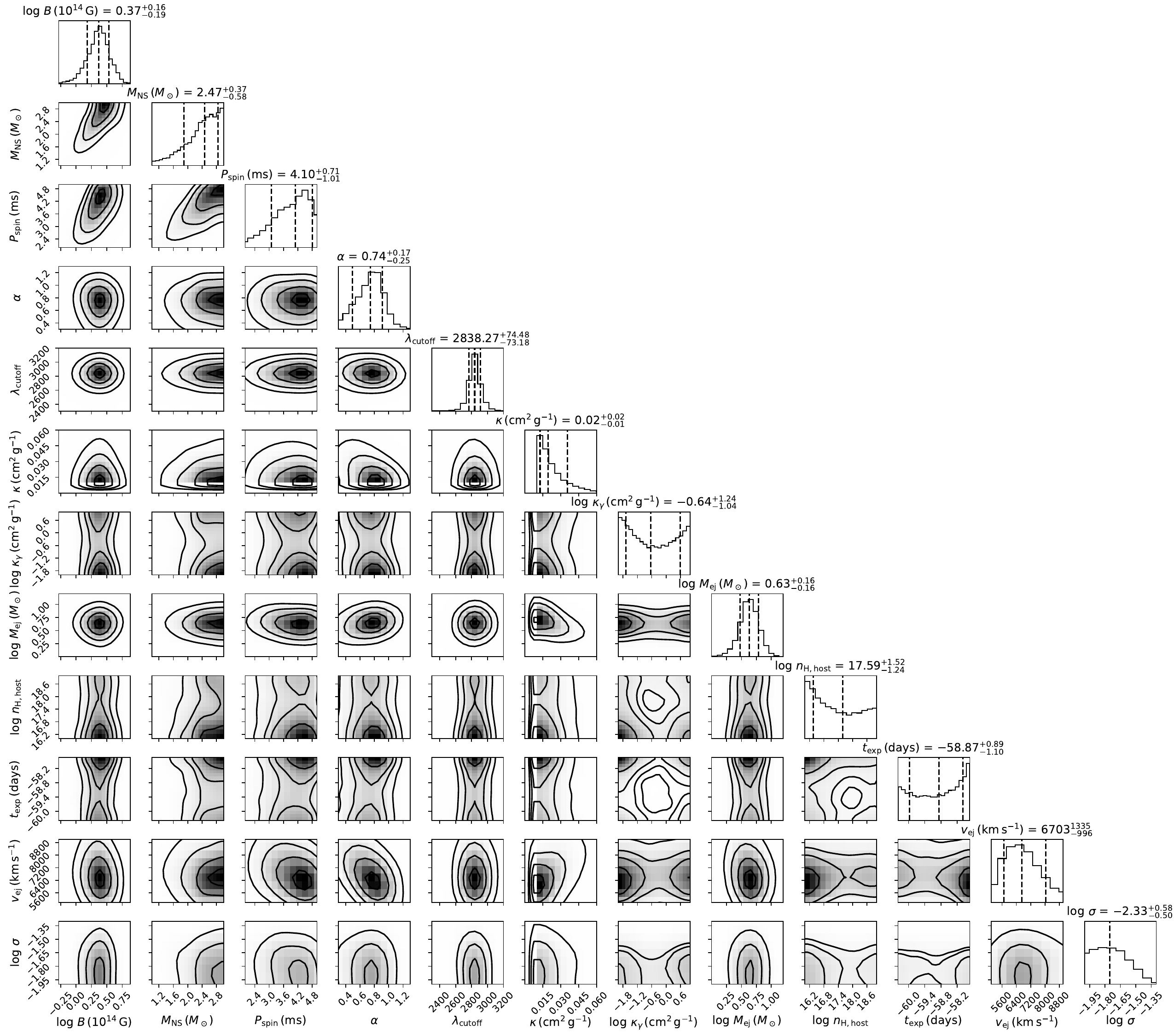}
        \caption{Corner plot of MOSFiT parameters fit to 2025wny.  The explosion time $t_{\rm exp}$ is given relative to the MJD of the first data point used in the fit, ${\rm MJD} = 60951.62$.}
        \label{fig:corner}
    \end{figure*}

\begin{appendix}
    Figure \ref{fig:corner} shows a corner plot of the full set of MOSFiT parameters for 2025wny.  For clarity, we omit the $f_{\mathrm{Ni}}$, $\theta_{\mathrm{PB}}$, and $T_{\rm min}$ parameters; these parameters are primarily constrained by the priors and do not appear strongly degenerate with other parameters.  Table \ref{tab:mosfit} lists the best-fit values and uncertainties for all parameters.

\end{appendix}

\end{document}